\documentclass[prl,twocolumn,10pt,aps,longbibliography,superscriptaddress]{revtex4-1}
\usepackage{amsmath,amssymb,bm}
\usepackage{graphicx}
\usepackage{epstopdf}
\usepackage{latexsym}
\usepackage{subfigure}
\usepackage{color}
\usepackage{natbib}
\usepackage{hyperref}
\usepackage{braket}
\usepackage{dsfont}
\usepackage[english]{babel}
\usepackage{blindtext}
\hypersetup{
  colorlinks,
  citecolor=magenta,
  linkcolor=blue,
  urlcolor=blue}
      
\begin{document}

\title{Universal features of entanglement entropy in the honeycomb Hubbard model}
\author{Jonathan D'Emidio}
\affiliation{Donostia International Physics Center, P. Manuel de Lardizabal 4, 20018 Donostia-San Sebasti\'an, Spain}
\author{Rom\'an Or\'us}
\affiliation{Donostia International Physics Center, P. Manuel de Lardizabal 4, 20018 Donostia-San Sebasti\'an, Spain}
\affiliation{IKERBASQUE, Basque Foundation for Science, Maria Diaz de Haro 3, 48013 Bilbao, Spain}
\affiliation{Multiverse Computing, Paseo de Miram\'on 170, E-20014 Donostia-San Sebasti\'an, Spain}
\author{Nicolas Laflorencie}
\affiliation{Laboratoire de Physique Th\'eorique, Universit\'e de Toulouse, CNRS, UPS, France}
\affiliation{Donostia International Physics Center, P. Manuel de Lardizabal 4, 20018 Donostia-San Sebasti\'an, Spain}
\author{Fernando de Juan}
\affiliation{Donostia International Physics Center, P. Manuel de Lardizabal 4, 20018 Donostia-San Sebasti\'an, Spain}
\affiliation{IKERBASQUE, Basque Foundation for Science, Maria Diaz de Haro 3, 48013 Bilbao, Spain}

\begin{abstract}
The entanglement entropy is a unique probe to reveal universal features of strongly interacting many-body systems.  In two or more dimensions these features are subtle, and detecting them numerically requires extreme precision, a notoriously difficult task.  This is especially challenging in models of interacting fermions, where many such universal features have yet to be observed.  In this paper we tackle this challenge by introducing a new method to compute the R\'enyi entanglement entropy in auxiliary-field quantum Monte Carlo simulations, where we treat the entangling region itself as a stochastic variable.  We demonstrate the efficiency of this method by extracting, for the first time, universal subleading logarithmic terms in a two dimensional model of interacting fermions, focusing on the half-filled honeycomb Hubbard model at $T=0$.  We detect the universal corner contribution due to gapless fermions throughout the Dirac semi-metal phase and at the Gross-Neveu-Yukawa critical point, where the latter shows a pronounced enhancement depending on the type of entangling cut.  Finally, we observe the universal Goldstone mode contribution in the antiferromagnetic Mott insulating phase.
\end{abstract}
\maketitle

The entanglement entropy (EE) quantifies the information shared between a subsystem and its environment in a quantum many-body wavefunction.  Remarkably, the EE finite-size scaling form has contributions that depend uniquely on universal physical quantities, making it a powerful probe to characterize strongly correlated systems.  A famous example of this is found in one-dimensional critical systems, where the EE grows logarithmically in the subsystem size with a prefactor given by the central charge [\onlinecite{Holzhey1994:Geometric, Vidal2003:EinQCP, Korepin2004:EE1D, Calabrese2004:EEinQFT}].  In two dimensions the EE grows in proportion to the boundary of the subsystem, the so-called ``area law" [\onlinecite{Eisert2010:AreaLaws}], but critical ground states display a subleading universal logarithmic contribution when the subsystem contains sharp corners [\onlinecite{Fradkin2006:QuantumDrum},\onlinecite{Casini2007:UniversalTerms}].  Additionally, in the case of continuous symmetry breaking, each Goldstone mode contributes a logarithmic term with a coefficient equal to one half [\onlinecite{Song2011:EEHeisen,Kallin2011:AnomaliesEEHeisen,Metlitski2011:GoldstoneEE}].  In the absence of symmetry breaking, topological states can be detected by a universal negative constant term in the EE [\onlinecite{Kitaev2006:TEE},\onlinecite{Levin2006:TEE}] as well as other entanglement measures [\onlinecite{Or_s_2014},\onlinecite{PhysRevLett.113.257202}].

Despite a wide variety of numerical work investigating spin and boson systems [\onlinecite{Laflorencie2016:Entanglement}], the universal features of EE of 2D interacting fermions have largely remained an unexplored frontier.  Important exceptions have focused on the universal constant in gapped systems, as in the case of the topological EE of fractional quantum hall ground states [\onlinecite{Zaletel2013:FQH},\onlinecite{Zhu2015:NAFQH}] and angle dependent constant of the $\nu=1/2$ Laughlin wavefunction [\onlinecite{Estienne2022:UniversalFluct}].  Additionally, the flux dependence of the constant term for gapless Dirac fermions was investigated in [\onlinecite{Zhu2018:EmergentDirac}].

Since the pioneering work of Grover [\onlinecite{Grover2013:EEinteracting}], auxiliary-field determinental quantum Monte Carlo (DQMC) simulations have offered a promising route to large-scale calculations of the R\'enyi EE of interacting fermions [\onlinecite{Grover2013:EEinteracting, Assaad2014:Espectra, Broecker2014:RenyiDQMC, Wang2014:REEcontinuous, Assaad2015:Stable, Drut2015:HybridQMCEE, Drut2016:Noise,Broecker2016:EESignProb,Broecker2016:StabilizationEE}]. However, the universal features of EE in 2D have remained out of reach for these methods. In the Grover method, one samples from uncorrelated replica configurations and the R\'enyi EE estimator suffers from rare events that dominate the statistical average. This problem becomes increasingly severe for larger entangling regions and interaction strengths.  More elaborate DQMC methods of computing the R\'enyi EE [\onlinecite{Broecker2014:RenyiDQMC},\onlinecite{Assaad2015:Stable}] offer better controlled statistical errors.  However, this comes at the price of increasing the effective number of degrees of freedom, making simulations more costly, and further requires special numerical stabilization techniques.  The lack of adequate techniques has even sparked interest in alternative probes of fermion entanglement that offer superior efficiency [\onlinecite{Jiang2022:FermionDisorderOp},\onlinecite{Liu2023:FermionDisorderGNYDQCP}].

In this work, we develop an improved method to compute the R\'enyi EE in DQMC that solves the above mentioned sampling problem and enables us to achieve unprecedented precision. To do so, we leverage recent advancements in computing the R\'enyi EE in quantum spin systems via nonequilibrium work [\onlinecite{Demidio2020:EEfromNEQLW, Block2020:KSL, Zhao2022:EEatDQC,Zhao2022:REEefficiency}], originally inspired by [\onlinecite{Alba2017:Jarzynski}], to develop an improved \textit{equilibrium} method for DQMC simulations. Our approach harnesses the power of importance sampling by introducing an extended ensemble of Monte Carlo configurations in which the entangling region is allowed to fluctuate.  Remarkably, the original formulation by Grover admits such an extended ensemble that can be simulated efficiently using standard DQMC techniques.  

\begin{figure}[!t]
\centerline{\includegraphics[angle=0,width=0.95\columnwidth]{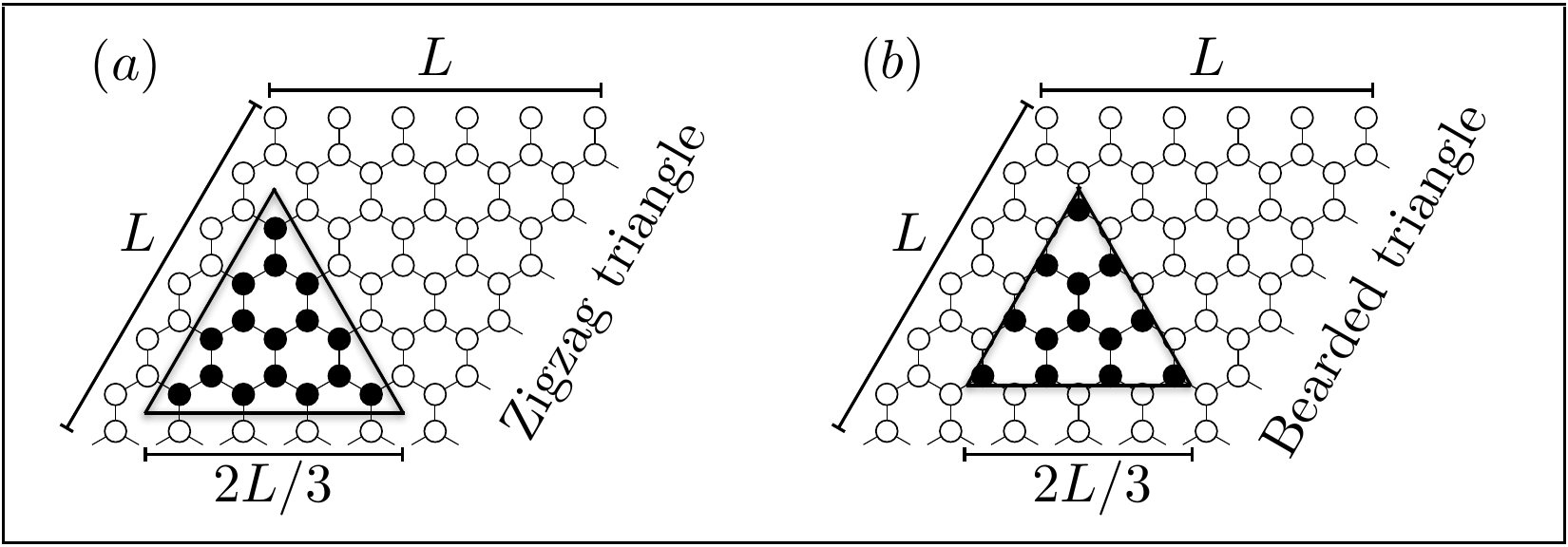}}
\caption{$(a)$ A triangular region on a $6 \times 6$ lattice with a zigzag edge.   $(b)$ A triangular region with a bearded edge.}
\label{fig:configs}
\end{figure}

In order to demonstrate the power of this technique, we use it to perform a comprehensive study of the logarithmic corrections to the area law in the half-filled honeycomb Hubbard model at $T=0$.  We first demonstrate the validity of our method by comparing to quasi-exact results obtained by density matrix renormalization group simulations (DMRG) [\onlinecite{itensor}].  Next we study the finite size scaling for two different triangular regions, see Fig. (\ref{fig:configs}), in the semi-metal phase and at the Gross-Neveu-Yukawa (GNY) critical point, where the latter shows an enhanced logarithmic contribution that depends on the type of entanglement cut.  This is further demonstrated by tracking the logarithmic contribution as a function of the interaction strength throughout the semi-metal phase and through the GNY point.  Finally, we compute the half-system R\'enyi EE at large interaction strength deep in the Mott insulating phase, revealing the distinct logarithmic contribution due to Goldstone modes.

{\em General method:}
We consider the general framework of auxiliary-field DQMC simulations, which map interacting fermionic systems to free fermions coupled to a fluctuating Hubbard-Stratonovich (HS) field [\onlinecite{Loh1992:DQMC},\onlinecite{Assaad2008:DQMC}].  For a given HS field configuration $s$, one has access to the equal-time Green's function $G^s_{ij} = \langle c_i c^{\dag}_j \rangle_s$.  In Grover's method [\onlinecite{Grover2013:EEinteracting}], the second R\'enyi EE, $S^{A}_2 =  - \ln \mathrm{Tr}(\rho^2_A)$, can be computed considering two independent replica DQMC simulations with Green's functions $G^{s_1}_{ij},  G^{s_2}_{ij}$ and taking the average:
\begin{equation}
\label{eq:eS2}
e^{-S^A_2} = \sum_{\{s_1\},\{s_2\}}P_{s_1}P_{s_2} \det \left( G^{s_1}_A G^{s_2}_A  + (\mathds{1}-G^{s_1}_A)(\mathds{1}-G^{s_2}_A)  \right).
\end{equation}
Here $G^{s}_A$ refers to the Green's function matrix that is restricted to the spatial region $A$, and $P_{s}$ is the probability of configuration $s$.  As previously mentioned, rare pairs of configurations $(s_1,s_2)$ give large contributions to Eq. (\ref{eq:eS2}).  To avoid this, we now show how to build correlations between the replica configurations such that the relevant phase space is better sampled, and in the process identify an improved Monte Carlo estimator for $S^A_2$.

Consider the distribution
\begin{equation}
\label{eq:ZA}
Z_A = \sum_{\{s_1\},\{s_2\}}W_{s_1}W_{s_2} \det g^{s_1,s_2}_{A},
\end{equation}
where we define the Grover matrix $g^{s_1,s_2}_{A} = G^{s_1}_A G^{s_2}_A  + (\mathds{1}-G^{s_1}_A)(\mathds{1}-G^{s_2}_A)$ and $W_s$ is the standard DQMC weight (unnormalized) of configuration $s$.  Eq. (\ref{eq:eS2}) can now be written as $e^{-S^A_2} =Z_A / Z_{\o}$, where $\o$ refers to the empty set, containing a zero dimensional Grover matrix with unit determinant.  
A highly efficient prescription for computing such partition function ratios was put forward in Ref. [\onlinecite{Demidio2020:EEfromNEQLW}].  Following this, we now consider a generalized ensemble made up of entangling regions $C$, which are proper subsets of the region $A$ [\onlinecite{suppmat}].  Furthermore, we control the distribution with an external field $\lambda$ that couples to the number of sites in the region $C$, denoted by $N_C$:
\begin{equation}
\label{eq:Zlam}
\mathcal{Z}(\lambda)=\sum_{C \subseteq A} \lambda^{N_C} (1-\lambda)^{N_A - N_C} Z_C, 
\end{equation}
where $Z_C$ is given by Eq. (\ref{eq:ZA}) with $A$ replaced by $C$. The ensemble in Eq. (\ref{eq:Zlam}) is designed such that $\mathcal{Z}(0)=Z_{\o}$ and $\mathcal{Z}(1)=Z_A$.  %The statistics of the stochastic variable $N_C$ can now be used to build an improved estimator for $S^A_2$.  At this point we interpret the $\lambda$ field as defining a certain energy for sites to be in the entangling region.  This allows us to develop a new \textit{equilibrium} estimator based on Zwanzig's formula for classical free energy differences [\onlinecite{Zwanzig1954:FEP}]
%\begin{equation}
%\label{eq:Zrat}
%\frac{\mathcal{Z}(\lambda_j)}{\mathcal{Z}(\lambda_i)}=\langle e^{E_{\lambda_i}(N_C) - E_{\lambda_j}(N_C) }\rangle_{\lambda_i}.
%\end{equation}
%Here $E_{\lambda}(N_C) = -N_C\ln(\lambda) - (N_A-N_C)\ln(1-\lambda)$, where dependence on the constant $N_A$ is omitted for ease of notation.  Since the distribution of $N_C$ at $\lambda=0,1$ is trivial, equilibrium simulations at intermediate values of $\lambda$ are performed and ratios can be computed using Eq. (\ref{eq:Zrat}), see Fig. (\ref{fig:configs}).  To make the best use of our available data, we opt to compute intermediate ratios using the two-sided Bennett acceptance ratio [\onlinecite{Bennett1976:BAR}], which has lower variance than other free energy estimators [\onlinecite{Shirts2005:CompareFEM}].

Given this, the ratio at two different values of $\lambda$ can be computed via a simple reweighting
\begin{equation}
\label{eq:Zrat}
\frac{\mathcal{Z}(\lambda_j)}{\mathcal{Z}(\lambda_i)}=\left\langle \left(\frac{\lambda_j}{\lambda_i}\right)^{N_C} \left(\frac{1-\lambda_j}{1-\lambda_i}\right)^{N_A-N_C} \right\rangle_{\lambda_i},
\end{equation}
where the only stochastic variable here is $N_C$ and the average is taken in the distribution according to $\lambda_i$.  In this way we may introduce many intermediate values of $\lambda$ in order to break up the overall exponentially small factor $\frac{Z_A} {Z_{\o}} = \frac{\mathcal{Z}(\lambda_1)}{\mathcal{Z}(0)}\frac{\mathcal{Z}(\lambda_2)}{\mathcal{Z}(\lambda_1)}...\frac{\mathcal{Z}(1)}{\mathcal{Z}(\lambda_{N_\lambda})}$ into computationally manageable pieces [\onlinecite{suppmat}].

A fundamental ingredient of our algorithm involves imbedding the Grover factor $\det(g^{s_1,s_2}_{C})$ directly into the DQMC configurational weight, as required to sample from the distribution in Eq. (\ref{eq:Zlam}). The inclusion of this factor is what effectively allows for importance sampling of the otherwise exponentially rare configurations appearing in Eq. (\ref{eq:eS2}). However, to include this factor in a manner that is both computationally efficient and numerically stable requires the resolution of a serious technical challenge, as we now describe.

Standard implementations of DQMC maintain an efficient computational complexity of $\mathcal{O}(N_{\tau} N^3_{\text{site}})$ by avoiding the explicit computation of fermionic determinants [\onlinecite{Loh1992:DQMC},\onlinecite{Assaad2008:DQMC}].  This technique relies on access to the equal-time Green's function $G^s_{ij}(\tau)$ located at the imaginary time slice that is being updated.  However, the Grover factor is always expressed in terms of Green's functions at a fixed imaginary time slice $G^s_{ij}(\theta)$, where observables are computed.  Na\"ively it would appear that the dependence of the configurational weight (including the Grover factor) on two different sets of Green's functions would render the standard fast update formulas inapplicable, making simulations prohibitively costly.  In the supplementary material [\onlinecite{suppmat}] we show how this crucial technical hurdle is overcome by making use of imaginary time displaced Green's functions [\onlinecite{Feldbacher2001:EfficientTimeDisp}], a standard object in most DQMC simulations.

{\em Model:}
As a benchmark system for our new method we select a classic model of interacting fermions in two dimensions: the Hubbard model on the honeycomb lattice at half filling. The Hamiltonian is given by
 \begin{equation}
\label{eq:hub}
H=-t\sum_{\langle i,j \rangle,\sigma} (c^{\dag}_{i,\sigma} c_{j,\sigma} + h. c.)  + U \sum_{i}(n_{i,\uparrow}-\tfrac{1}{2})(n_{i,\downarrow}-\tfrac{1}{2}).
\end{equation}
The restriction to half filling ensures the absence a sign problem, which equally applies to our generalized ensemble in Eq. (\ref{eq:Zlam}).  This model is known to host a semi-metal phase with a gapless Dirac spectrum up to $U_c \approx 3.8$ [\onlinecite{Sorella2012:Absence},\onlinecite{Assaad2013:Pinning}], beyond which the system enters a Mott insulating antiferromagnetically ordered phase with Goldstone modes from the spontaneous breaking of spin rotation symmetry.  The critical point at $U_c=3.8$ is in the GNY chiral Heisenberg universality class [\onlinecite{Herbut2006:Honeycomb}].

In the gapless semi-metal (SM) phase it is known that the R\'enyi EE of a triangular region with three sharp corners, depicted in Fig. (\ref{fig:configs}$a$) and (\ref{fig:configs}$b$), scales according to [\onlinecite{Fradkin2006:QuantumDrum},\onlinecite{Casini2007:UniversalTerms}]
\begin{equation}
\label{eq:gaplessEE}
S_2 = \mathcal{A}L - 3 a^{\mathrm{SM}}_2(\pi/3)\ln(L) + \text{const}. 
\end{equation}
Here $ a^{\mathrm{SM}}_2(\pi/3) \approx 0.1324$ [\onlinecite{Helmes2016:UniversalCorner}] is the universal coefficient from one $\pi/3$ corner with four free Dirac fermions (two spin and two valley).  A similar scaling form is also expected to hold at the GNY critical point, albeit with an \textit{unknown} value for the corner coefficient.  We point out that unbiased numerical simulations are the only means of estimating universal corner contributions at interacting fixed points, as was previously done for the 2+1d Ising, XY, and Heisenberg universality classes [\onlinecite{Singh2012:ThermoSingEE,Kallin2013:EE2DQCP,Kallin2014:CornerO3,Helmes2014:EEbilayer,Stoudenmire2014:CornerEEO2,Helmes2015:EEXY}].  These studies support the notion of the corner coefficient as a measure of the number of effective low-energy degrees of freedom [\onlinecite{Bueno2015:UniversalCorner},\onlinecite{Bueno2015:CornerContributions}].  We are therefore interested in comparing the value at the GNY point to that of free Dirac fermions.

In the Mott insulating phase, the contribution from Goldstone modes to the R\'enyi EE for a smooth entangling cut, as depicted in Fig. (\ref{fig:goldstone}), has a similar form [\onlinecite{Metlitski2011:GoldstoneEE}]:
\begin{equation}
\label{eq:goldstoneEE}
S_2 = \mathcal{A}L + \frac{N_g}{2}\ln(L) + \text{const}. 
\end{equation}
Here the logarithmic piece counts the number of Goldstone modes $N_g$, and comes with the opposite sign.  Since the honeycomb Hubbard model exhibits a known corner term in the semi-metal phase, an unknown corner term at the critical point, and an expected contribution from $N_g=2$ Goldstone modes in the Mott insulating phase, it serves as the perfect testbed to extract universal logs for the first time with our technique.

\begin{figure}[!t]
\centerline{\includegraphics[angle=0,width=1.0\columnwidth]{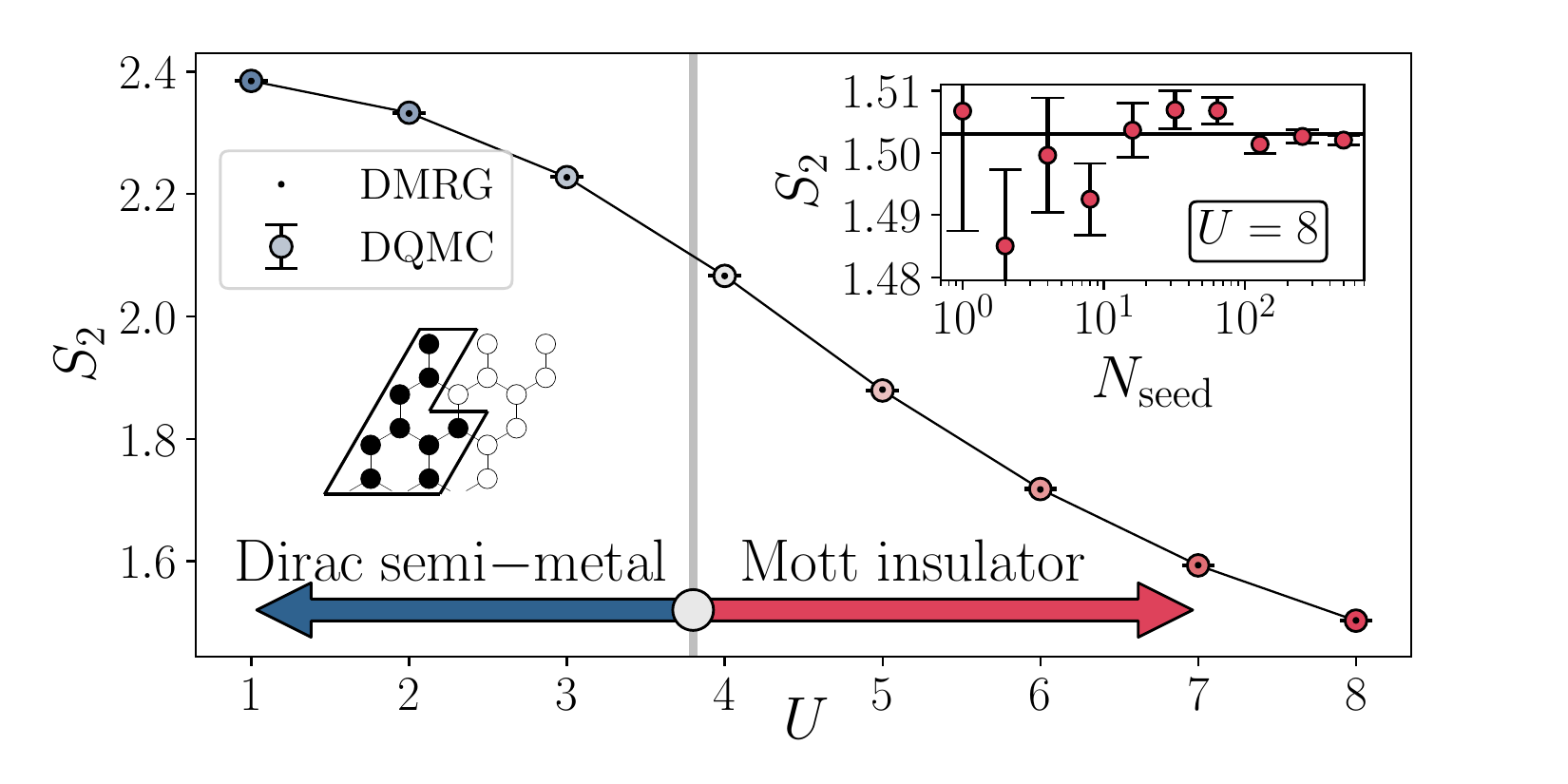}}
\caption{The new DQMC computation of $S_2$ as compared to DMRG [\onlinecite{itensor}] for the system pictured.  Here we use open boundaries in the x-direction.  The inset shows a closer look at the precise agreement at $U=8$ using only a single value of $\lambda=0.5$ and $N_{\mathrm{seed}}$ independent simulations.}
\label{fig:qmcvsed}
\end{figure}

%\begin{figure}[!t]
%\centerline{\includegraphics[angle=0,width=1.0\columnwidth]{U1andfreefig.pdf}}
%\caption{$S_2$ of the triangular geometry depicted in Fig. (\ref{fig:configs}$d$) at $U=1$.  Inset ($a$): data obtained using the same geometry with $U=0$ up to system size $L=60$.  Inset ($b$):  Fit to the corner coefficient in Eq. (\ref{eq:gaplessEE}) using a fit window [$L_{\text{min}}$,$L_{\max}$] for both the DQMC and free fermion data.   We find good agreement with the field theoretic value $a^{\mathrm{SM}}_2(\pi/3)\approx 0.1324$ [\onlinecite{Helmes2016:UniversalCorner}].}
%\label{fig:U1triangle}
%\end{figure}

{\em DQMC results:}
We implemented the $T=0$ projector DQMC algorithm (though our method can also be extended to finite temperature) with a symmetric Trotter decomposition and an SU(2) invariant HS transformation [\onlinecite{Loh1992:DQMC},\onlinecite{Assaad2008:DQMC},\onlinecite{Assaad1999:SU2Invariant}]. The calculations presented here used a Trotter step of $\Delta_{\tau}=0.1$ unless otherwise noted [\onlinecite{suppmat}].

We begin by comparing our method to quasi-exact results obtained by the DMRG method [\onlinecite{itensor}] on an $L = 3$ open cylinder as a function of $U$, shown in Fig. (\ref{fig:qmcvsed}).  Here we use a small Trotter step of $\Delta_{\tau}=0.01$ and a ground state projection of $\theta=10$ (2000 total Trotter slices).  We find perfect agreement with the DQMC results, and already at this system size we can see the qualitative feature of the semi-metal to Mott insulator transition near $U_c = 3.8$.  We note that on larger system sizes more values of $\lambda$ are needed such that adjacent distributions of $N_C$ have good overlap.  Details of the $\lambda$ values used in this work are given in the supplementary material [\onlinecite{suppmat}].

\begin{figure}[!h]
\centerline{\includegraphics[angle=0,width=1.0\columnwidth]{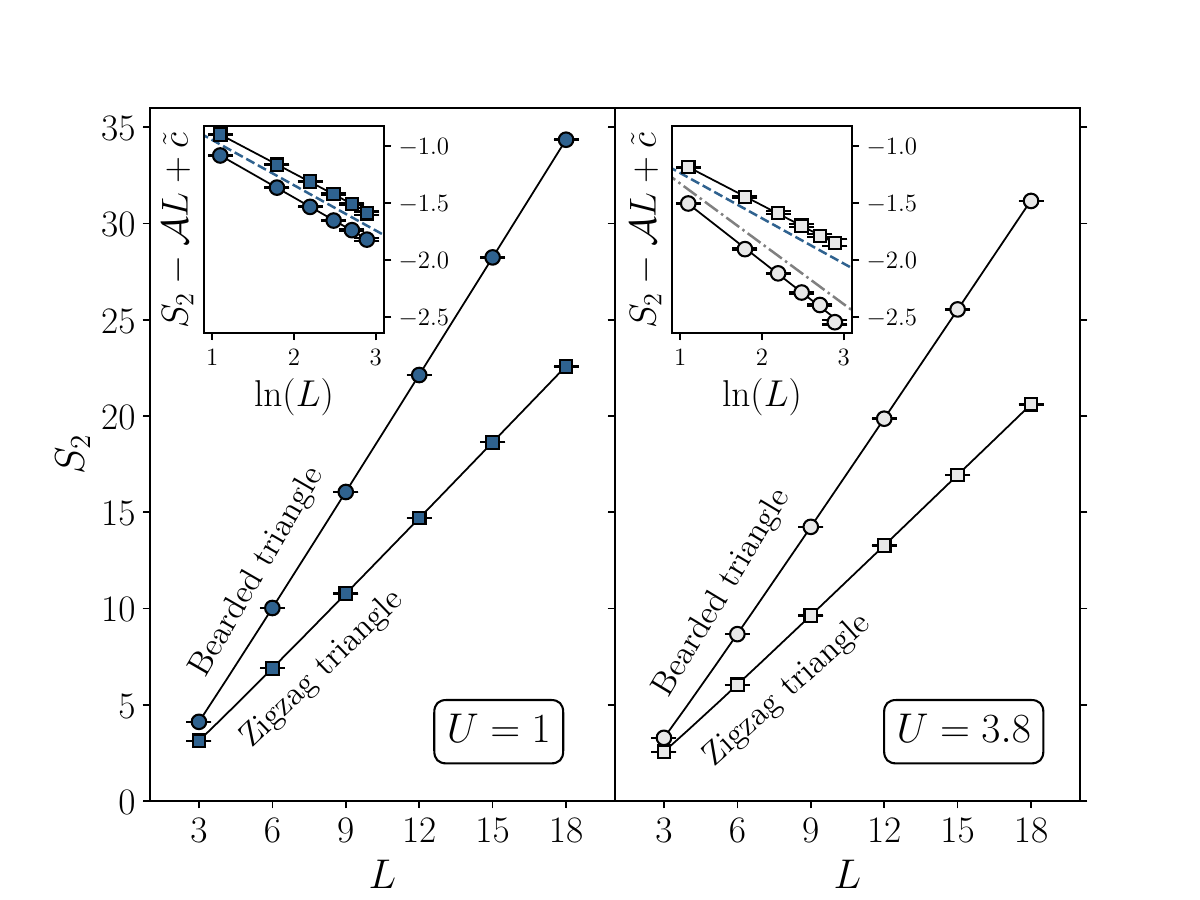}}
\caption{Left panel: $S_2$ computed for both kinds of triangular regions depicted in Fig. (\ref{fig:configs}$a$) and (\ref{fig:configs}$b$) at $U=1$ and twisted boundary conditions with $\phi=0.1$.  The inset shows the result of a three parameter fit to a linear plus log scaling and subtracting away the area law piece (a shift of $\tilde{c}=1$ is given to the bearded triangle for clarity of comparison).  The blue dashed line shows the field theory prediction for the semi-metal phase.  Right panel:  the same analysis but for with $U=3.8$ at the GNY point with $\phi=0$.  The bearded triangle shows an enhanced logarithmic contribution in this case.  This is compared to the field theory value for free fermions plus a three component boson shown by the grey dotted-dashed line in the inset.}
\label{fig:U3p8triangle}
\end{figure}

\begin{figure}[!h]
\centerline{\includegraphics[angle=0,width=1.0\columnwidth]{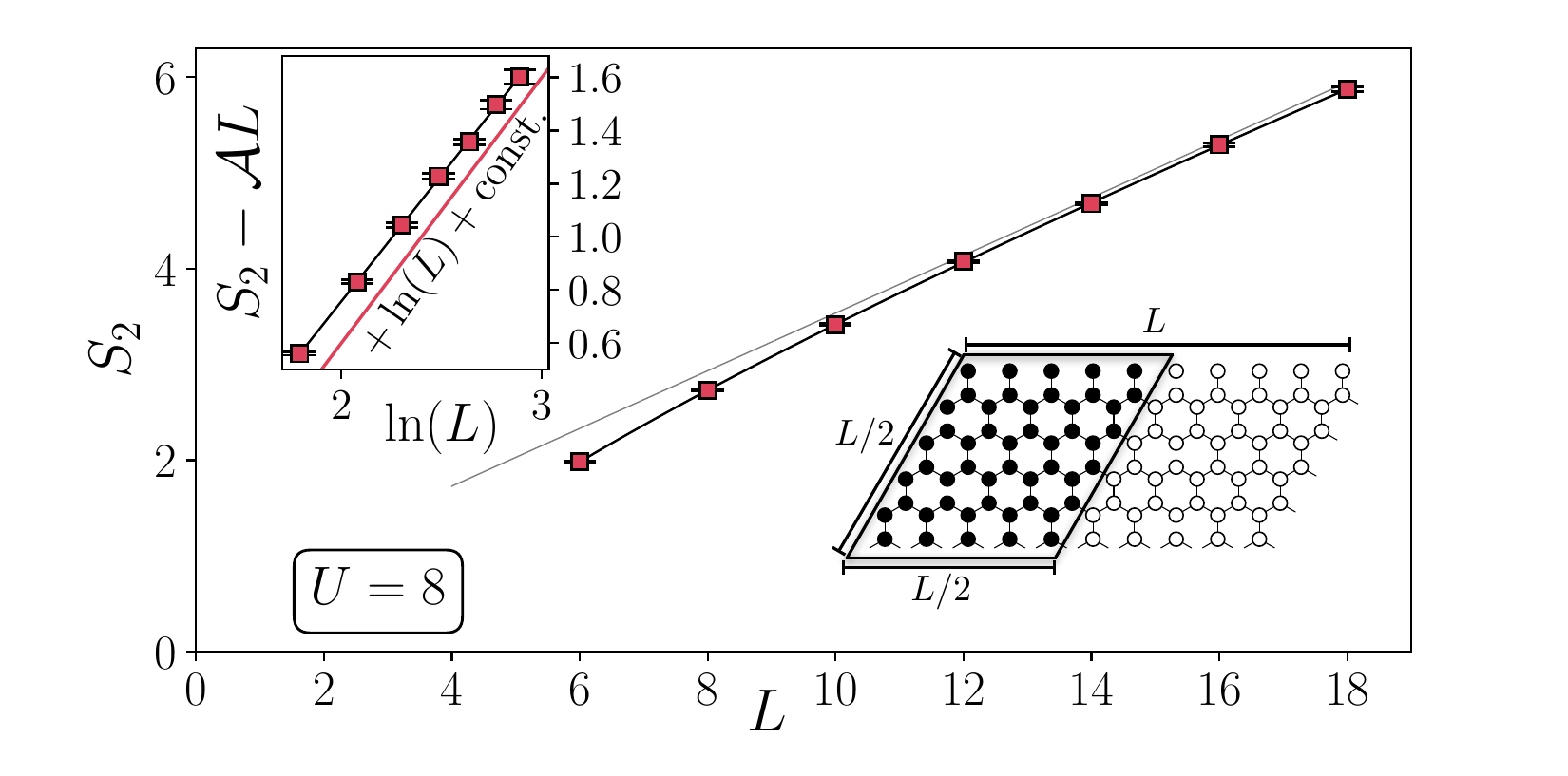}}
\caption{Half-system entropy $S_2$ on rectangular systems as depicted, with $U=8$.  This plot is similar to Fig. (\ref{fig:U3p8triangle}), but now the subleading log term has the opposite sign, as can be seen by the slope in the inset.  The thin line in the main panel helps to visualize the bend in the data coming from the logarithmic term.  The coefficient of the log term counts the Goldstone modes with the coefficient $N_g/2$, here giving $0.95(5)$.}
\label{fig:goldstone}
\end{figure}

We now move on to confirm the expected behavior at $U=1$.  In the left panel of Fig. (\ref{fig:U3p8triangle}) we show DQMC results using the two different triangular regions depicted in Fig. (\ref{fig:configs}$a$) and (\ref{fig:configs}$b$).  We have found that the ground state convergence of $S_2$ in the semi-metal phase is heavily affected by the proximity to the Dirac point.  We therefore use twisted boundary conditions in the $x$-direction, where hoppings that wrap the $x$-boundary get multiplied by the phase $e^{i 2\pi \phi}$ with $\phi=0.1$ in order to shift the Dirac point [\onlinecite{suppmat}].  We use projection times up to $\theta=2L$ on our largest system sizes to ensure convergence.  We find that both triangles give subleading logarithmic contributions that are consistent with the field theoretic value $a^{\mathrm{SM}}_2(\pi/3)\approx0.1324$ [\onlinecite{Helmes2016:UniversalCorner}], namely we find $a^{\mathrm{SM}}_2(\pi/3)=0.126(4), 0.136(2)$ for the zigzag and bearded triangles, respectively.  This is shown in the inset, where the fit to the area law term is subtracted away and the result is plotted versus $\ln(L)$.

\begin{figure}[t]
\centerline{\includegraphics[angle=0,width=1.0\columnwidth]{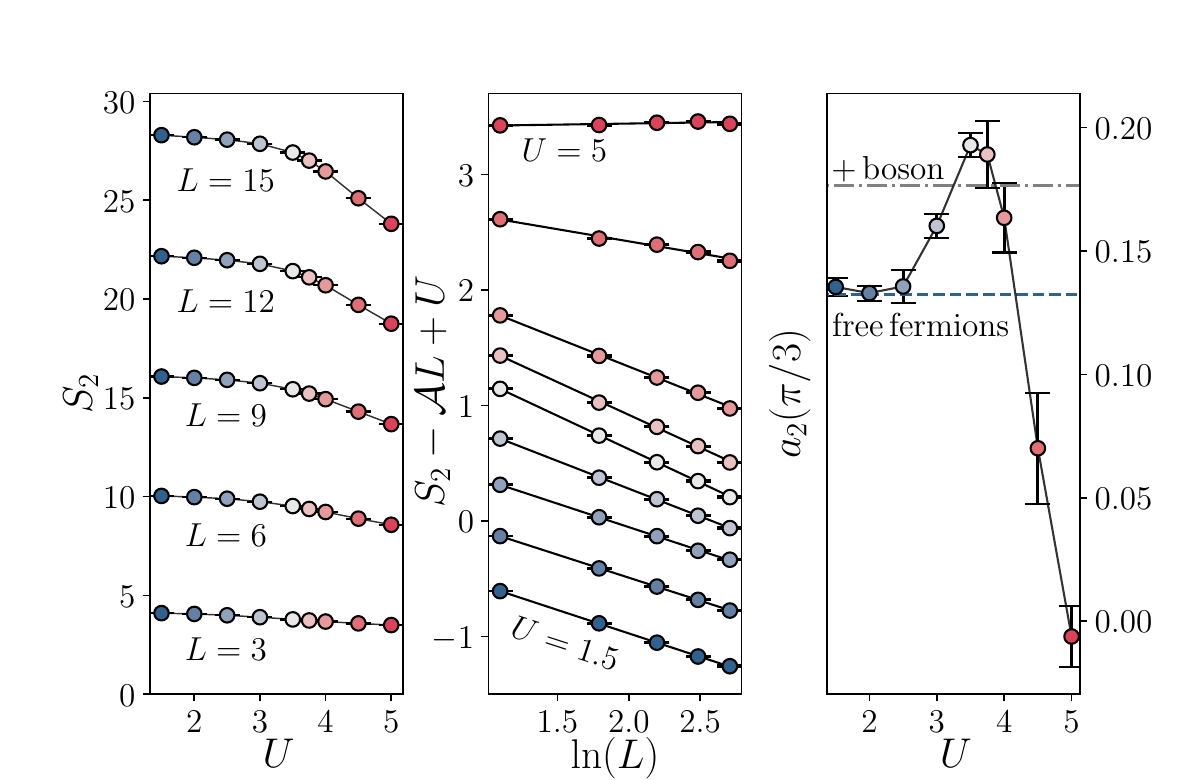}}
\caption{Left panel: $S_2$ for bearded triangles depicted in Fig. (\ref{fig:configs}$b$) using $\Delta_{\tau}=0.5$ and $\phi=0.15$ as a function of $U$.  Middle panel:  The fit as a function of $L$ performed for each value of $U$ with the area law piece subtracted away, with a constant shift of $U$ added for clarity.  Right panel:  The extracted corner coefficient as a function of $U$, showing an enhanced value at the GNY point.}
\label{fig:cornerVsU}
\end{figure}

Next in the right panel of Fig. (\ref{fig:U3p8triangle}) we perform the same analysis but with $U=3.8$, at the GNY point.  Here we find that ground state convergence is easier than in the semi-metal phase and so we set $\phi=0$ but still use $\theta = 2 L$ on the largest system sizes. We interestingly find a difference in the logarithmic contributions between the zigzag and bearded triangles.  The zigzag triangle gives a similar value to free Dirac fermions: $a^{\mathrm{GNY}}_2(\pi/3) = 0.124(5)$, while the bearded triangle shows an enhanced logarithmic contribution with $a^{\mathrm{GNY}}_2(\pi/3) =0.187(12)$. It is appropriate to compare this with what is expected from free Dirac fermions plus a three component gapless boson ($\mathrm{f + b}$) representing the N\'eel order parameter.  The field theoretic value in this case gives $a^{\mathrm{f+b}}_2(\pi/3) \approx 0.1764$ [\onlinecite{Helmes2016:UniversalCorner}], in the same range as the bearded triangle.  While it is difficult to estimate the true value at the GNY point, which we expect to be less than the previously quoted field theory value [\onlinecite{wwkprivate}], the fact that our finite size data produces a comparable value is encouraging and motivates us to investigate this logarithmic term in more detail.

We wish to now study the bearded triangle corner coefficient in detail as a function of $U$.  In order to do this we use a much larger Trotter time step $\Delta_{\tau}=0.5$, since we have found the logarithmic terms to be independent of of the Trotter step [\onlinecite{suppmat}].  We further fix $\theta = L$ (except $L=3$ where we use $\theta = 2L$) and $\phi = 0.15$.  The left panel of Fig. (\ref{fig:cornerVsU}) shows the resulting $S_2$ data as a function of $U$.  For each value of $U$ we perform a fit as a function of $L$, then in the middle panel we reveal the logarithmic contribution by subtracting off the fitted area law piece.  The resulting slopes are plotted in the rightmost panel, where we see clear agreement with the free fermion result throughout the semi-metal phase with a pronounced enhancement at the GNY point followed by a sharp drop into the Mott insulating phase.  The values at the GNY point are consistent with Fig. (\ref{fig:U3p8triangle}) even though here we have used twisted boundary conditions and a much larger Trotter step.

Finally, we further increase $U$ into the antiferromagnetic Mott insulator phase where we expect to see a \textit{positive} subleading logarithmic contribution coming from Goldstone modes, given by Eq. (\ref{eq:goldstoneEE}).  Fig. (\ref{fig:goldstone}) shows the half-system entropy on rectangular systems as shown in the figure with $U=8$.  We see a clear positive log with a coefficient in agreement with the contribution from two Goldstone modes. 

{\em Conclusions:}
We introduced an equilibrium Monte Carlo estimator for the R\'enyi EE in interacting fermion systems that allows for importance sampling of the original estimator by Grover.  We used this method to detect, for the first time, logarithmic corrections to the area law in a 2D model of interacting fermions.  Importantly, we find that such logarithmic terms can be sensitive to the type of entanglement cut that is used.  

{\em Note added:}
Since our method first appeared, it has been independently implemented and applied to several different fermionic models [\onlinecite{DaLiao:2023pdn,Pan:2023ysg,Liu:2023msa}].  Additionally, building on our methodology, an even more efficient protocol was developed in [\onlinecite{DaLiao:2023klr}] that obviates the need to sample over subset entangling regions.

{\em Acknowledgements:} We gratefully acknowledge William Witczak-Krempa for stimulating correspondence.  R. O. and F. J. acknowledge funding from the Spanish MCI/AEI/FEDER (grant PGC2018-101988-B-C21).  Technical and human support provided by DIPC Supercomputing Center, where the main results were collected, is gratefully acknowledged.  Additional resources were used from the XSEDE allocation NSF DMR-130040 using the Expanse cluster at SDSC.

\bibliography{HubbardEE}

\begin{thebibliography}{58}%
\makeatletter
\providecommand \@ifxundefined [1]{%
 \@ifx{#1\undefined}
}%
\providecommand \@ifnum [1]{%
 \ifnum #1\expandafter \@firstoftwo
 \else \expandafter \@secondoftwo
 \fi
}%
\providecommand \@ifx [1]{%
 \ifx #1\expandafter \@firstoftwo
 \else \expandafter \@secondoftwo
 \fi
}%
\providecommand \natexlab [1]{#1}%
\providecommand \enquote  [1]{``#1''}%
\providecommand \bibnamefont  [1]{#1}%
\providecommand \bibfnamefont [1]{#1}%
\providecommand \citenamefont [1]{#1}%
\providecommand \href@noop [0]{\@secondoftwo}%
\providecommand \href [0]{\begingroup \@sanitize@url \@href}%
\providecommand \@href[1]{\@@startlink{#1}\@@href}%
\providecommand \@@href[1]{\endgroup#1\@@endlink}%
\providecommand \@sanitize@url [0]{\catcode `\\12\catcode `\$12\catcode
  `\&12\catcode `\#12\catcode `\^12\catcode `\_12\catcode `\%12\relax}%
\providecommand \@@startlink[1]{}%
\providecommand \@@endlink[0]{}%
\providecommand \url  [0]{\begingroup\@sanitize@url \@url }%
\providecommand \@url [1]{\endgroup\@href {#1}{\urlprefix }}%
\providecommand \urlprefix  [0]{URL }%
\providecommand \Eprint [0]{\href }%
\providecommand \doibase [0]{http://dx.doi.org/}%
\providecommand \selectlanguage [0]{\@gobble}%
\providecommand \bibinfo  [0]{\@secondoftwo}%
\providecommand \bibfield  [0]{\@secondoftwo}%
\providecommand \translation [1]{[#1]}%
\providecommand \BibitemOpen [0]{}%
\providecommand \bibitemStop [0]{}%
\providecommand \bibitemNoStop [0]{.\EOS\space}%
\providecommand \EOS [0]{\spacefactor3000\relax}%
\providecommand \BibitemShut  [1]{\csname bibitem#1\endcsname}%
\let\auto@bib@innerbib\@empty
%</preamble>
\bibitem [{\citenamefont {Holzhey}\ \emph {et~al.}(1994)\citenamefont
  {Holzhey}, \citenamefont {Larsen},\ and\ \citenamefont
  {Wilczek}}]{Holzhey1994:Geometric}%
  \BibitemOpen
  \bibfield  {author} {\bibinfo {author} {\bibfnamefont {Christoph}\
  \bibnamefont {Holzhey}}, \bibinfo {author} {\bibfnamefont {Finn}\
  \bibnamefont {Larsen}}, \ and\ \bibinfo {author} {\bibfnamefont {Frank}\
  \bibnamefont {Wilczek}},\ }\bibfield  {title} {\enquote {\bibinfo {title}
  {Geometric and renormalized entropy in conformal field theory},}\ }\href
  {\doibase https://doi.org/10.1016/0550-3213(94)90402-2} {\bibfield  {journal}
  {\bibinfo  {journal} {Nuclear Physics B}\ }\textbf {\bibinfo {volume}
  {424}},\ \bibinfo {pages} {443 -- 467} (\bibinfo {year} {1994})}\BibitemShut
  {NoStop}%
\bibitem [{\citenamefont {Vidal}\ \emph {et~al.}(2003)\citenamefont {Vidal},
  \citenamefont {Latorre}, \citenamefont {Rico},\ and\ \citenamefont
  {Kitaev}}]{Vidal2003:EinQCP}%
  \BibitemOpen
  \bibfield  {author} {\bibinfo {author} {\bibfnamefont {G.}~\bibnamefont
  {Vidal}}, \bibinfo {author} {\bibfnamefont {J.~I.}\ \bibnamefont {Latorre}},
  \bibinfo {author} {\bibfnamefont {E.}~\bibnamefont {Rico}}, \ and\ \bibinfo
  {author} {\bibfnamefont {A.}~\bibnamefont {Kitaev}},\ }\bibfield  {title}
  {\enquote {\bibinfo {title} {Entanglement in quantum critical phenomena},}\
  }\href {\doibase 10.1103/PhysRevLett.90.227902} {\bibfield  {journal}
  {\bibinfo  {journal} {Phys. Rev. Lett.}\ }\textbf {\bibinfo {volume} {90}},\
  \bibinfo {pages} {227902} (\bibinfo {year} {2003})}\BibitemShut {NoStop}%
\bibitem [{\citenamefont {Korepin}(2004)}]{Korepin2004:EE1D}%
  \BibitemOpen
  \bibfield  {author} {\bibinfo {author} {\bibfnamefont {V.~E.}\ \bibnamefont
  {Korepin}},\ }\bibfield  {title} {\enquote {\bibinfo {title} {Universality of
  entropy scaling in one dimensional gapless models},}\ }\href {\doibase
  10.1103/PhysRevLett.92.096402} {\bibfield  {journal} {\bibinfo  {journal}
  {Phys. Rev. Lett.}\ }\textbf {\bibinfo {volume} {92}},\ \bibinfo {pages}
  {096402} (\bibinfo {year} {2004})}\BibitemShut {NoStop}%
\bibitem [{\citenamefont {Calabrese}\ and\ \citenamefont
  {Cardy}(2004)}]{Calabrese2004:EEinQFT}%
  \BibitemOpen
  \bibfield  {author} {\bibinfo {author} {\bibfnamefont {Pasquale}\
  \bibnamefont {Calabrese}}\ and\ \bibinfo {author} {\bibfnamefont {John}\
  \bibnamefont {Cardy}},\ }\bibfield  {title} {\enquote {\bibinfo {title}
  {Entanglement entropy and quantum field theory},}\ }\href {\doibase
  10.1088/1742-5468/2004/06/p06002} {\bibfield  {journal} {\bibinfo  {journal}
  {Journal of Statistical Mechanics: Theory and Experiment}\ }\textbf {\bibinfo
  {volume} {2004}},\ \bibinfo {pages} {P06002} (\bibinfo {year}
  {2004})}\BibitemShut {NoStop}%
\bibitem [{\citenamefont {Eisert}\ \emph {et~al.}(2010)\citenamefont {Eisert},
  \citenamefont {Cramer},\ and\ \citenamefont {Plenio}}]{Eisert2010:AreaLaws}%
  \BibitemOpen
  \bibfield  {author} {\bibinfo {author} {\bibfnamefont {J.}~\bibnamefont
  {Eisert}}, \bibinfo {author} {\bibfnamefont {M.}~\bibnamefont {Cramer}}, \
  and\ \bibinfo {author} {\bibfnamefont {M.~B.}\ \bibnamefont {Plenio}},\
  }\bibfield  {title} {\enquote {\bibinfo {title} {Colloquium: Area laws for
  the entanglement entropy},}\ }\href {\doibase 10.1103/RevModPhys.82.277}
  {\bibfield  {journal} {\bibinfo  {journal} {Rev. Mod. Phys.}\ }\textbf
  {\bibinfo {volume} {82}},\ \bibinfo {pages} {277--306} (\bibinfo {year}
  {2010})}\BibitemShut {NoStop}%
\bibitem [{\citenamefont {Fradkin}\ and\ \citenamefont
  {Moore}(2006)}]{Fradkin2006:QuantumDrum}%
  \BibitemOpen
  \bibfield  {author} {\bibinfo {author} {\bibfnamefont {Eduardo}\ \bibnamefont
  {Fradkin}}\ and\ \bibinfo {author} {\bibfnamefont {Joel~E.}\ \bibnamefont
  {Moore}},\ }\bibfield  {title} {\enquote {\bibinfo {title} {Entanglement
  entropy of 2d conformal quantum critical points: Hearing the shape of a
  quantum drum},}\ }\href {\doibase 10.1103/PhysRevLett.97.050404} {\bibfield
  {journal} {\bibinfo  {journal} {Phys. Rev. Lett.}\ }\textbf {\bibinfo
  {volume} {97}},\ \bibinfo {pages} {050404} (\bibinfo {year}
  {2006})}\BibitemShut {NoStop}%
\bibitem [{\citenamefont {Casini}\ and\ \citenamefont
  {Huerta}(2007)}]{Casini2007:UniversalTerms}%
  \BibitemOpen
  \bibfield  {author} {\bibinfo {author} {\bibfnamefont {H.}~\bibnamefont
  {Casini}}\ and\ \bibinfo {author} {\bibfnamefont {M.}~\bibnamefont
  {Huerta}},\ }\bibfield  {title} {\enquote {\bibinfo {title} {Universal terms
  for the entanglement entropy in 2+1 dimensions},}\ }\href {\doibase
  https://doi.org/10.1016/j.nuclphysb.2006.12.012} {\bibfield  {journal}
  {\bibinfo  {journal} {Nuclear Physics B}\ }\textbf {\bibinfo {volume}
  {764}},\ \bibinfo {pages} {183--201} (\bibinfo {year} {2007})}\BibitemShut
  {NoStop}%
\bibitem [{\citenamefont {Song}\ \emph {et~al.}(2011)\citenamefont {Song},
  \citenamefont {Laflorencie}, \citenamefont {Rachel},\ and\ \citenamefont
  {Le~Hur}}]{Song2011:EEHeisen}%
  \BibitemOpen
  \bibfield  {author} {\bibinfo {author} {\bibfnamefont {H.~Francis}\
  \bibnamefont {Song}}, \bibinfo {author} {\bibfnamefont {Nicolas}\
  \bibnamefont {Laflorencie}}, \bibinfo {author} {\bibfnamefont {Stephan}\
  \bibnamefont {Rachel}}, \ and\ \bibinfo {author} {\bibfnamefont {Karyn}\
  \bibnamefont {Le~Hur}},\ }\bibfield  {title} {\enquote {\bibinfo {title}
  {Entanglement entropy of the two-dimensional heisenberg antiferromagnet},}\
  }\href {\doibase 10.1103/PhysRevB.83.224410} {\bibfield  {journal} {\bibinfo
  {journal} {Phys. Rev. B}\ }\textbf {\bibinfo {volume} {83}},\ \bibinfo
  {pages} {224410} (\bibinfo {year} {2011})}\BibitemShut {NoStop}%
\bibitem [{\citenamefont {Kallin}\ \emph {et~al.}(2011)\citenamefont {Kallin},
  \citenamefont {Hastings}, \citenamefont {Melko},\ and\ \citenamefont
  {Singh}}]{Kallin2011:AnomaliesEEHeisen}%
  \BibitemOpen
  \bibfield  {author} {\bibinfo {author} {\bibfnamefont {Ann~B.}\ \bibnamefont
  {Kallin}}, \bibinfo {author} {\bibfnamefont {Matthew~B.}\ \bibnamefont
  {Hastings}}, \bibinfo {author} {\bibfnamefont {Roger~G.}\ \bibnamefont
  {Melko}}, \ and\ \bibinfo {author} {\bibfnamefont {Rajiv R.~P.}\ \bibnamefont
  {Singh}},\ }\bibfield  {title} {\enquote {\bibinfo {title} {Anomalies in the
  entanglement properties of the square-lattice heisenberg model},}\ }\href
  {\doibase 10.1103/PhysRevB.84.165134} {\bibfield  {journal} {\bibinfo
  {journal} {Phys. Rev. B}\ }\textbf {\bibinfo {volume} {84}},\ \bibinfo
  {pages} {165134} (\bibinfo {year} {2011})}\BibitemShut {NoStop}%
\bibitem [{\citenamefont {{Metlitski}}\ and\ \citenamefont
  {{Grover}}(2011)}]{Metlitski2011:GoldstoneEE}%
  \BibitemOpen
  \bibfield  {author} {\bibinfo {author} {\bibfnamefont {Max~A.}\ \bibnamefont
  {{Metlitski}}}\ and\ \bibinfo {author} {\bibfnamefont {Tarun}\ \bibnamefont
  {{Grover}}},\ }\bibfield  {title} {\enquote {\bibinfo {title} {{Entanglement
  Entropy of Systems with Spontaneously Broken Continuous Symmetry}},}\
  }\href@noop {} {\bibfield  {journal} {\bibinfo  {journal} {arXiv e-prints}\
  ,\ \bibinfo {eid} {arXiv:1112.5166}} (\bibinfo {year} {2011})},\ \Eprint
  {http://arxiv.org/abs/1112.5166} {arXiv:1112.5166 [cond-mat.str-el]}
  \BibitemShut {NoStop}%
\bibitem [{\citenamefont {Kitaev}\ and\ \citenamefont
  {Preskill}(2006)}]{Kitaev2006:TEE}%
  \BibitemOpen
  \bibfield  {author} {\bibinfo {author} {\bibfnamefont {Alexei}\ \bibnamefont
  {Kitaev}}\ and\ \bibinfo {author} {\bibfnamefont {John}\ \bibnamefont
  {Preskill}},\ }\bibfield  {title} {\enquote {\bibinfo {title} {Topological
  entanglement entropy},}\ }\href {\doibase 10.1103/PhysRevLett.96.110404}
  {\bibfield  {journal} {\bibinfo  {journal} {Phys. Rev. Lett.}\ }\textbf
  {\bibinfo {volume} {96}},\ \bibinfo {pages} {110404} (\bibinfo {year}
  {2006})}\BibitemShut {NoStop}%
\bibitem [{\citenamefont {Levin}\ and\ \citenamefont
  {Wen}(2006)}]{Levin2006:TEE}%
  \BibitemOpen
  \bibfield  {author} {\bibinfo {author} {\bibfnamefont {Michael}\ \bibnamefont
  {Levin}}\ and\ \bibinfo {author} {\bibfnamefont {Xiao-Gang}\ \bibnamefont
  {Wen}},\ }\bibfield  {title} {\enquote {\bibinfo {title} {Detecting
  topological order in a ground state wave function},}\ }\href {\doibase
  10.1103/PhysRevLett.96.110405} {\bibfield  {journal} {\bibinfo  {journal}
  {Phys. Rev. Lett.}\ }\textbf {\bibinfo {volume} {96}},\ \bibinfo {pages}
  {110405} (\bibinfo {year} {2006})}\BibitemShut {NoStop}%
\bibitem [{\citenamefont {Or{\'{u}}s}\ \emph {et~al.}(2014)\citenamefont
  {Or{\'{u}}s}, \citenamefont {Wei}, \citenamefont {Buerschaper},\ and\
  \citenamefont {den Nest}}]{Or_s_2014}%
  \BibitemOpen
  \bibfield  {author} {\bibinfo {author} {\bibfnamefont {Rom{\'{a}}n}\
  \bibnamefont {Or{\'{u}}s}}, \bibinfo {author} {\bibfnamefont {Tzu-Chieh}\
  \bibnamefont {Wei}}, \bibinfo {author} {\bibfnamefont {Oliver}\ \bibnamefont
  {Buerschaper}}, \ and\ \bibinfo {author} {\bibfnamefont {Maarten~Van}\
  \bibnamefont {den Nest}},\ }\bibfield  {title} {\enquote {\bibinfo {title}
  {Geometric entanglement in topologically ordered states},}\ }\href {\doibase
  10.1088/1367-2630/16/1/013015} {\bibfield  {journal} {\bibinfo  {journal}
  {New Journal of Physics}\ }\textbf {\bibinfo {volume} {16}},\ \bibinfo
  {pages} {013015} (\bibinfo {year} {2014})}\BibitemShut {NoStop}%
\bibitem [{\citenamefont {Or\'us}\ \emph {et~al.}(2014)\citenamefont {Or\'us},
  \citenamefont {Wei}, \citenamefont {Buerschaper},\ and\ \citenamefont
  {Garc\'{\i}a-Saez}}]{PhysRevLett.113.257202}%
  \BibitemOpen
  \bibfield  {author} {\bibinfo {author} {\bibfnamefont {Rom\'an}\ \bibnamefont
  {Or\'us}}, \bibinfo {author} {\bibfnamefont {Tzu-Chieh}\ \bibnamefont {Wei}},
  \bibinfo {author} {\bibfnamefont {Oliver}\ \bibnamefont {Buerschaper}}, \
  and\ \bibinfo {author} {\bibfnamefont {Artur}\ \bibnamefont
  {Garc\'{\i}a-Saez}},\ }\bibfield  {title} {\enquote {\bibinfo {title}
  {Topological transitions from multipartite entanglement with tensor networks:
  A procedure for sharper and faster characterization},}\ }\href {\doibase
  10.1103/PhysRevLett.113.257202} {\bibfield  {journal} {\bibinfo  {journal}
  {Phys. Rev. Lett.}\ }\textbf {\bibinfo {volume} {113}},\ \bibinfo {pages}
  {257202} (\bibinfo {year} {2014})}\BibitemShut {NoStop}%
\bibitem [{\citenamefont {Laflorencie}(2016)}]{Laflorencie2016:Entanglement}%
  \BibitemOpen
  \bibfield  {author} {\bibinfo {author} {\bibfnamefont {Nicolas}\ \bibnamefont
  {Laflorencie}},\ }\bibfield  {title} {\enquote {\bibinfo {title} {Quantum
  entanglement in condensed matter systems},}\ }\href {\doibase
  https://doi.org/10.1016/j.physrep.2016.06.008} {\bibfield  {journal}
  {\bibinfo  {journal} {Physics Reports}\ }\textbf {\bibinfo {volume} {646}},\
  \bibinfo {pages} {1--59} (\bibinfo {year} {2016})}\BibitemShut {NoStop}%
\bibitem [{\citenamefont {Zaletel}\ \emph {et~al.}(2013)\citenamefont
  {Zaletel}, \citenamefont {Mong},\ and\ \citenamefont
  {Pollmann}}]{Zaletel2013:FQH}%
  \BibitemOpen
  \bibfield  {author} {\bibinfo {author} {\bibfnamefont {Michael~P.}\
  \bibnamefont {Zaletel}}, \bibinfo {author} {\bibfnamefont {Roger S.~K.}\
  \bibnamefont {Mong}}, \ and\ \bibinfo {author} {\bibfnamefont {Frank}\
  \bibnamefont {Pollmann}},\ }\bibfield  {title} {\enquote {\bibinfo {title}
  {Topological characterization of fractional quantum hall ground states from
  microscopic hamiltonians},}\ }\href {\doibase 10.1103/PhysRevLett.110.236801}
  {\bibfield  {journal} {\bibinfo  {journal} {Phys. Rev. Lett.}\ }\textbf
  {\bibinfo {volume} {110}},\ \bibinfo {pages} {236801} (\bibinfo {year}
  {2013})}\BibitemShut {NoStop}%
\bibitem [{\citenamefont {Zhu}\ \emph {et~al.}(2015)\citenamefont {Zhu},
  \citenamefont {Gong}, \citenamefont {Haldane},\ and\ \citenamefont
  {Sheng}}]{Zhu2015:NAFQH}%
  \BibitemOpen
  \bibfield  {author} {\bibinfo {author} {\bibfnamefont {W.}~\bibnamefont
  {Zhu}}, \bibinfo {author} {\bibfnamefont {S.~S.}\ \bibnamefont {Gong}},
  \bibinfo {author} {\bibfnamefont {F.~D.~M.}\ \bibnamefont {Haldane}}, \ and\
  \bibinfo {author} {\bibfnamefont {D.~N.}\ \bibnamefont {Sheng}},\ }\bibfield
  {title} {\enquote {\bibinfo {title} {Fractional quantum hall states at
  $\ensuremath{\nu}=13/5$ and $12/5$ and their non-abelian nature},}\ }\href
  {\doibase 10.1103/PhysRevLett.115.126805} {\bibfield  {journal} {\bibinfo
  {journal} {Phys. Rev. Lett.}\ }\textbf {\bibinfo {volume} {115}},\ \bibinfo
  {pages} {126805} (\bibinfo {year} {2015})}\BibitemShut {NoStop}%
\bibitem [{\citenamefont {Estienne}\ \emph {et~al.}(2022)\citenamefont
  {Estienne}, \citenamefont {St{\'e}phan},\ and\ \citenamefont
  {Witczak-Krempa}}]{Estienne2022:UniversalFluct}%
  \BibitemOpen
  \bibfield  {author} {\bibinfo {author} {\bibfnamefont {Benoit}\ \bibnamefont
  {Estienne}}, \bibinfo {author} {\bibfnamefont {Jean-Marie}\ \bibnamefont
  {St{\'e}phan}}, \ and\ \bibinfo {author} {\bibfnamefont {William}\
  \bibnamefont {Witczak-Krempa}},\ }\bibfield  {title} {\enquote {\bibinfo
  {title} {Cornering the universal shape of fluctuations},}\ }\href {\doibase
  10.1038/s41467-021-27727-1} {\bibfield  {journal} {\bibinfo  {journal}
  {Nature Communications}\ }\textbf {\bibinfo {volume} {13}},\ \bibinfo {pages}
  {287} (\bibinfo {year} {2022})}\BibitemShut {NoStop}%
\bibitem [{\citenamefont {Zhu}\ \emph {et~al.}(2018)\citenamefont {Zhu},
  \citenamefont {Chen}, \citenamefont {He},\ and\ \citenamefont
  {Witczak-Krempa}}]{Zhu2018:EmergentDirac}%
  \BibitemOpen
  \bibfield  {author} {\bibinfo {author} {\bibfnamefont {Wei}\ \bibnamefont
  {Zhu}}, \bibinfo {author} {\bibfnamefont {Xiao}\ \bibnamefont {Chen}},
  \bibinfo {author} {\bibfnamefont {Yin-Chen}\ \bibnamefont {He}}, \ and\
  \bibinfo {author} {\bibfnamefont {William}\ \bibnamefont {Witczak-Krempa}},\
  }\bibfield  {title} {\enquote {\bibinfo {title} {Entanglement signatures of
  emergent dirac fermions: Kagome spin liquid and quantum criticality},}\
  }\href {\doibase 10.1126/sciadv.aat5535} {\bibfield  {journal} {\bibinfo
  {journal} {Science Advances}\ }\textbf {\bibinfo {volume} {4}},\ \bibinfo
  {pages} {eaat5535} (\bibinfo {year} {2018})},\ \Eprint
  {http://arxiv.org/abs/https://www.science.org/doi/pdf/10.1126/sciadv.aat5535}
  {https://www.science.org/doi/pdf/10.1126/sciadv.aat5535} \BibitemShut
  {NoStop}%
\bibitem [{\citenamefont {Grover}(2013)}]{Grover2013:EEinteracting}%
  \BibitemOpen
  \bibfield  {author} {\bibinfo {author} {\bibfnamefont {Tarun}\ \bibnamefont
  {Grover}},\ }\bibfield  {title} {\enquote {\bibinfo {title} {Entanglement of
  interacting fermions in quantum monte carlo calculations},}\ }\href {\doibase
  10.1103/PhysRevLett.111.130402} {\bibfield  {journal} {\bibinfo  {journal}
  {Phys. Rev. Lett.}\ }\textbf {\bibinfo {volume} {111}},\ \bibinfo {pages}
  {130402} (\bibinfo {year} {2013})}\BibitemShut {NoStop}%
\bibitem [{\citenamefont {Assaad}\ \emph {et~al.}(2014)\citenamefont {Assaad},
  \citenamefont {Lang},\ and\ \citenamefont {Toldin}}]{Assaad2014:Espectra}%
  \BibitemOpen
  \bibfield  {author} {\bibinfo {author} {\bibfnamefont {F.F.}\ \bibnamefont
  {Assaad}}, \bibinfo {author} {\bibfnamefont {T.C.}\ \bibnamefont {Lang}}, \
  and\ \bibinfo {author} {\bibfnamefont {Francesco~Parisen}\ \bibnamefont
  {Toldin}},\ }\bibfield  {title} {\enquote {\bibinfo {title} {Entanglement
  spectra of interacting fermions in quantum monte carlo simulations},}\ }\href
  {\doibase 10.1103/PhysRevB.89.125121} {\bibfield  {journal} {\bibinfo
  {journal} {Phys. Rev. B}\ }\textbf {\bibinfo {volume} {89}},\ \bibinfo
  {pages} {125121} (\bibinfo {year} {2014})}\BibitemShut {NoStop}%
\bibitem [{\citenamefont {Broecker}\ and\ \citenamefont
  {Trebst}(2014)}]{Broecker2014:RenyiDQMC}%
  \BibitemOpen
  \bibfield  {author} {\bibinfo {author} {\bibfnamefont {Peter}\ \bibnamefont
  {Broecker}}\ and\ \bibinfo {author} {\bibfnamefont {Simon}\ \bibnamefont
  {Trebst}},\ }\bibfield  {title} {\enquote {\bibinfo {title} {R{\'{e}}nyi
  entropies of interacting fermions from determinantal quantum monte carlo
  simulations},}\ }\href@noop {} {\bibfield  {journal} {\bibinfo  {journal}
  {Journal of Statistical Mechanics: Theory and Experiment}\ }\textbf {\bibinfo
  {volume} {2014}},\ \bibinfo {pages} {P08015} (\bibinfo {year}
  {2014})}\BibitemShut {NoStop}%
\bibitem [{\citenamefont {Wang}\ and\ \citenamefont
  {Troyer}(2014)}]{Wang2014:REEcontinuous}%
  \BibitemOpen
  \bibfield  {author} {\bibinfo {author} {\bibfnamefont {Lei}\ \bibnamefont
  {Wang}}\ and\ \bibinfo {author} {\bibfnamefont {Matthias}\ \bibnamefont
  {Troyer}},\ }\bibfield  {title} {\enquote {\bibinfo {title} {Renyi
  entanglement entropy of interacting fermions calculated using the
  continuous-time quantum monte carlo method},}\ }\href {\doibase
  10.1103/PhysRevLett.113.110401} {\bibfield  {journal} {\bibinfo  {journal}
  {Phys. Rev. Lett.}\ }\textbf {\bibinfo {volume} {113}},\ \bibinfo {pages}
  {110401} (\bibinfo {year} {2014})}\BibitemShut {NoStop}%
\bibitem [{\citenamefont {Assaad}(2015)}]{Assaad2015:Stable}%
  \BibitemOpen
  \bibfield  {author} {\bibinfo {author} {\bibfnamefont {Fakher~F.}\
  \bibnamefont {Assaad}},\ }\bibfield  {title} {\enquote {\bibinfo {title}
  {Stable quantum monte carlo simulations for entanglement spectra of
  interacting fermions},}\ }\href {\doibase 10.1103/PhysRevB.91.125146}
  {\bibfield  {journal} {\bibinfo  {journal} {Phys. Rev. B}\ }\textbf {\bibinfo
  {volume} {91}},\ \bibinfo {pages} {125146} (\bibinfo {year}
  {2015})}\BibitemShut {NoStop}%
\bibitem [{\citenamefont {Drut}\ and\ \citenamefont
  {Porter}(2015)}]{Drut2015:HybridQMCEE}%
  \BibitemOpen
  \bibfield  {author} {\bibinfo {author} {\bibfnamefont {Joaqu\'{\i}n~E.}\
  \bibnamefont {Drut}}\ and\ \bibinfo {author} {\bibfnamefont {William~J.}\
  \bibnamefont {Porter}},\ }\bibfield  {title} {\enquote {\bibinfo {title}
  {Hybrid monte carlo approach to the entanglement entropy of interacting
  fermions},}\ }\href {\doibase 10.1103/PhysRevB.92.125126} {\bibfield
  {journal} {\bibinfo  {journal} {Phys. Rev. B}\ }\textbf {\bibinfo {volume}
  {92}},\ \bibinfo {pages} {125126} (\bibinfo {year} {2015})}\BibitemShut
  {NoStop}%
\bibitem [{\citenamefont {Drut}\ and\ \citenamefont
  {Porter}(2016)}]{Drut2016:Noise}%
  \BibitemOpen
  \bibfield  {author} {\bibinfo {author} {\bibfnamefont {Joaqu\'{\i}n~E.}\
  \bibnamefont {Drut}}\ and\ \bibinfo {author} {\bibfnamefont {William~J.}\
  \bibnamefont {Porter}},\ }\bibfield  {title} {\enquote {\bibinfo {title}
  {Entanglement, noise, and the cumulant expansion},}\ }\href {\doibase
  10.1103/PhysRevE.93.043301} {\bibfield  {journal} {\bibinfo  {journal} {Phys.
  Rev. E}\ }\textbf {\bibinfo {volume} {93}},\ \bibinfo {pages} {043301}
  (\bibinfo {year} {2016})}\BibitemShut {NoStop}%
\bibitem [{\citenamefont {Broecker}\ and\ \citenamefont
  {Trebst}(2016{\natexlab{a}})}]{Broecker2016:EESignProb}%
  \BibitemOpen
  \bibfield  {author} {\bibinfo {author} {\bibfnamefont {Peter}\ \bibnamefont
  {Broecker}}\ and\ \bibinfo {author} {\bibfnamefont {Simon}\ \bibnamefont
  {Trebst}},\ }\bibfield  {title} {\enquote {\bibinfo {title} {Entanglement and
  the fermion sign problem in auxiliary field quantum monte carlo
  simulations},}\ }\href {\doibase 10.1103/PhysRevB.94.075144} {\bibfield
  {journal} {\bibinfo  {journal} {Phys. Rev. B}\ }\textbf {\bibinfo {volume}
  {94}},\ \bibinfo {pages} {075144} (\bibinfo {year}
  {2016}{\natexlab{a}})}\BibitemShut {NoStop}%
\bibitem [{\citenamefont {Broecker}\ and\ \citenamefont
  {Trebst}(2016{\natexlab{b}})}]{Broecker2016:StabilizationEE}%
  \BibitemOpen
  \bibfield  {author} {\bibinfo {author} {\bibfnamefont {Peter}\ \bibnamefont
  {Broecker}}\ and\ \bibinfo {author} {\bibfnamefont {Simon}\ \bibnamefont
  {Trebst}},\ }\bibfield  {title} {\enquote {\bibinfo {title} {Numerical
  stabilization of entanglement computation in auxiliary-field quantum monte
  carlo simulations of interacting many-fermion systems},}\ }\href {\doibase
  10.1103/PhysRevE.94.063306} {\bibfield  {journal} {\bibinfo  {journal} {Phys.
  Rev. E}\ }\textbf {\bibinfo {volume} {94}},\ \bibinfo {pages} {063306}
  (\bibinfo {year} {2016}{\natexlab{b}})}\BibitemShut {NoStop}%
\bibitem [{\citenamefont {Jiang}\ \emph {et~al.}(2023)\citenamefont {Jiang},
  \citenamefont {Chen}, \citenamefont {Liu}, \citenamefont {Rong},
  \citenamefont {Assaad}, \citenamefont {Cheng}, \citenamefont {Sun},\ and\
  \citenamefont {Meng}}]{Jiang2022:FermionDisorderOp}%
  \BibitemOpen
  \bibfield  {author} {\bibinfo {author} {\bibfnamefont {Weilun}\ \bibnamefont
  {Jiang}}, \bibinfo {author} {\bibfnamefont {Bin-Bin}\ \bibnamefont {Chen}},
  \bibinfo {author} {\bibfnamefont {Zi~Hong}\ \bibnamefont {Liu}}, \bibinfo
  {author} {\bibfnamefont {Junchen}\ \bibnamefont {Rong}}, \bibinfo {author}
  {\bibfnamefont {Fakher~F.}\ \bibnamefont {Assaad}}, \bibinfo {author}
  {\bibfnamefont {Meng}\ \bibnamefont {Cheng}}, \bibinfo {author}
  {\bibfnamefont {Kai}\ \bibnamefont {Sun}}, \ and\ \bibinfo {author}
  {\bibfnamefont {Zi~Yang}\ \bibnamefont {Meng}},\ }\bibfield  {title}
  {\enquote {\bibinfo {title} {Many versus one: The disorder operator and
  entanglement entropy in fermionic quantum matter},}\ }\href {\doibase
  10.21468/scipostphys.15.3.082} {\bibfield  {journal} {\bibinfo  {journal}
  {SciPost Physics}\ }\textbf {\bibinfo {volume} {15}} (\bibinfo {year}
  {2023}),\ 10.21468/scipostphys.15.3.082}\BibitemShut {NoStop}%
\bibitem [{\citenamefont {Liu}\ \emph {et~al.}(2023{\natexlab{a}})\citenamefont
  {Liu}, \citenamefont {Jiang}, \citenamefont {Chen}, \citenamefont {Rong},
  \citenamefont {Cheng}, \citenamefont {Sun}, \citenamefont {Meng},\ and\
  \citenamefont {Assaad}}]{Liu2023:FermionDisorderGNYDQCP}%
  \BibitemOpen
  \bibfield  {author} {\bibinfo {author} {\bibfnamefont {Zi~Hong}\ \bibnamefont
  {Liu}}, \bibinfo {author} {\bibfnamefont {Weilun}\ \bibnamefont {Jiang}},
  \bibinfo {author} {\bibfnamefont {Bin-Bin}\ \bibnamefont {Chen}}, \bibinfo
  {author} {\bibfnamefont {Junchen}\ \bibnamefont {Rong}}, \bibinfo {author}
  {\bibfnamefont {Meng}\ \bibnamefont {Cheng}}, \bibinfo {author}
  {\bibfnamefont {Kai}\ \bibnamefont {Sun}}, \bibinfo {author} {\bibfnamefont
  {Zi~Yang}\ \bibnamefont {Meng}}, \ and\ \bibinfo {author} {\bibfnamefont
  {Fakher~F.}\ \bibnamefont {Assaad}},\ }\bibfield  {title} {\enquote {\bibinfo
  {title} {Fermion disorder operator at gross-neveu and deconfined quantum
  criticalities},}\ }\href {\doibase 10.1103/PhysRevLett.130.266501} {\bibfield
   {journal} {\bibinfo  {journal} {Phys. Rev. Lett.}\ }\textbf {\bibinfo
  {volume} {130}},\ \bibinfo {pages} {266501} (\bibinfo {year}
  {2023}{\natexlab{a}})}\BibitemShut {NoStop}%
\bibitem [{\citenamefont {D'Emidio}(2020)}]{Demidio2020:EEfromNEQLW}%
  \BibitemOpen
  \bibfield  {author} {\bibinfo {author} {\bibfnamefont {Jonathan}\
  \bibnamefont {D'Emidio}},\ }\bibfield  {title} {\enquote {\bibinfo {title}
  {Entanglement entropy from nonequilibrium work},}\ }\href {\doibase
  10.1103/PhysRevLett.124.110602} {\bibfield  {journal} {\bibinfo  {journal}
  {Phys. Rev. Lett.}\ }\textbf {\bibinfo {volume} {124}},\ \bibinfo {pages}
  {110602} (\bibinfo {year} {2020})}\BibitemShut {NoStop}%
\bibitem [{\citenamefont {Block}\ \emph {et~al.}(2020)\citenamefont {Block},
  \citenamefont {D'Emidio},\ and\ \citenamefont {Kaul}}]{Block2020:KSL}%
  \BibitemOpen
  \bibfield  {author} {\bibinfo {author} {\bibfnamefont {M.S.}\ \bibnamefont
  {Block}}, \bibinfo {author} {\bibfnamefont {Jonathan}\ \bibnamefont
  {D'Emidio}}, \ and\ \bibinfo {author} {\bibfnamefont {R.~K.}\ \bibnamefont
  {Kaul}},\ }\bibfield  {title} {\enquote {\bibinfo {title} {Kagome model for a
  {$\mathbb{Z}_2$} quantum spin liquid},}\ }\href {\doibase
  10.1103/PhysRevB.101.020402} {\bibfield  {journal} {\bibinfo  {journal}
  {Phys. Rev. B}\ }\textbf {\bibinfo {volume} {101}},\ \bibinfo {pages}
  {020402} (\bibinfo {year} {2020})}\BibitemShut {NoStop}%
\bibitem [{\citenamefont {Zhao}\ \emph
  {et~al.}(2022{\natexlab{a}})\citenamefont {Zhao}, \citenamefont {Wang},
  \citenamefont {Yan}, \citenamefont {Cheng},\ and\ \citenamefont
  {Meng}}]{Zhao2022:EEatDQC}%
  \BibitemOpen
  \bibfield  {author} {\bibinfo {author} {\bibfnamefont {Jiarui}\ \bibnamefont
  {Zhao}}, \bibinfo {author} {\bibfnamefont {Yan-Cheng}\ \bibnamefont {Wang}},
  \bibinfo {author} {\bibfnamefont {Zheng}\ \bibnamefont {Yan}}, \bibinfo
  {author} {\bibfnamefont {Meng}\ \bibnamefont {Cheng}}, \ and\ \bibinfo
  {author} {\bibfnamefont {Zi~Yang}\ \bibnamefont {Meng}},\ }\bibfield  {title}
  {\enquote {\bibinfo {title} {Scaling of entanglement entropy at deconfined
  quantum criticality},}\ }\href {\doibase 10.1103/PhysRevLett.128.010601}
  {\bibfield  {journal} {\bibinfo  {journal} {Phys. Rev. Lett.}\ }\textbf
  {\bibinfo {volume} {128}},\ \bibinfo {pages} {010601} (\bibinfo {year}
  {2022}{\natexlab{a}})}\BibitemShut {NoStop}%
\bibitem [{\citenamefont {Zhao}\ \emph
  {et~al.}(2022{\natexlab{b}})\citenamefont {Zhao}, \citenamefont {Chen},
  \citenamefont {Wang}, \citenamefont {Yan}, \citenamefont {Cheng},\ and\
  \citenamefont {Meng}}]{Zhao2022:REEefficiency}%
  \BibitemOpen
  \bibfield  {author} {\bibinfo {author} {\bibfnamefont {Jiarui}\ \bibnamefont
  {Zhao}}, \bibinfo {author} {\bibfnamefont {Bin-Bin}\ \bibnamefont {Chen}},
  \bibinfo {author} {\bibfnamefont {Yan-Cheng}\ \bibnamefont {Wang}}, \bibinfo
  {author} {\bibfnamefont {Zheng}\ \bibnamefont {Yan}}, \bibinfo {author}
  {\bibfnamefont {Meng}\ \bibnamefont {Cheng}}, \ and\ \bibinfo {author}
  {\bibfnamefont {Zi~Yang}\ \bibnamefont {Meng}},\ }\bibfield  {title}
  {\enquote {\bibinfo {title} {Measuring r{\'e}nyi entanglement entropy with
  high efficiency and precision in quantum monte carlo simulations},}\
  }\href@noop {} {\bibfield  {journal} {\bibinfo  {journal} {npj Quantum
  Materials}\ }\textbf {\bibinfo {volume} {7}},\ \bibinfo {pages} {69}
  (\bibinfo {year} {2022}{\natexlab{b}})}\BibitemShut {NoStop}%
\bibitem [{\citenamefont {Alba}(2017)}]{Alba2017:Jarzynski}%
  \BibitemOpen
  \bibfield  {author} {\bibinfo {author} {\bibfnamefont {Vincenzo}\
  \bibnamefont {Alba}},\ }\bibfield  {title} {\enquote {\bibinfo {title}
  {Out-of-equilibrium protocol for r\'{e}nyi entropies via the jarzynski
  equality},}\ }\href {\doibase 10.1103/PhysRevE.95.062132} {\bibfield
  {journal} {\bibinfo  {journal} {Phys. Rev. E}\ }\textbf {\bibinfo {volume}
  {95}},\ \bibinfo {pages} {062132} (\bibinfo {year} {2017})}\BibitemShut
  {NoStop}%
\bibitem [{\citenamefont {Fishman}\ \emph {et~al.}(2020)\citenamefont
  {Fishman}, \citenamefont {White},\ and\ \citenamefont
  {Stoudenmire}}]{itensor}%
  \BibitemOpen
  \bibfield  {author} {\bibinfo {author} {\bibfnamefont {Matthew}\ \bibnamefont
  {Fishman}}, \bibinfo {author} {\bibfnamefont {Steven~R.}\ \bibnamefont
  {White}}, \ and\ \bibinfo {author} {\bibfnamefont {E.~Miles}\ \bibnamefont
  {Stoudenmire}},\ }\href@noop {} {\enquote {\bibinfo {title} {The
  \mbox{ITensor} software library for tensor network calculations},}\ }
  (\bibinfo {year} {2020}),\ \Eprint {http://arxiv.org/abs/2007.14822}
  {arXiv:2007.14822} \BibitemShut {NoStop}%
\bibitem [{\citenamefont {Loh}\ and\ \citenamefont
  {Gubernatis}(1992)}]{Loh1992:DQMC}%
  \BibitemOpen
  \bibfield  {author} {\bibinfo {author} {\bibfnamefont {E.Y.}\ \bibnamefont
  {Loh}}\ and\ \bibinfo {author} {\bibfnamefont {J.E.}\ \bibnamefont
  {Gubernatis}},\ }\bibfield  {title} {\enquote {\bibinfo {title} {Chapter 4 -
  stable numerical simulations of models of interacting electrons in
  condensed-matter physics},}\ }in\ \href {\doibase
  https://doi.org/10.1016/B978-0-444-88885-3.50009-3} {\emph {\bibinfo
  {booktitle} {Electronic Phase Transitions}}},\ \bibinfo {series} {Modern
  Problems in Condensed Matter Sciences}, Vol.~\bibinfo {volume} {32},\
  \bibinfo {editor} {edited by\ \bibinfo {editor} {\bibfnamefont
  {W.}~\bibnamefont {Hanke}}\ and\ \bibinfo {editor} {\bibfnamefont {Yu.V.}\
  \bibnamefont {Kopaev}}}\ (\bibinfo  {publisher} {Elsevier},\ \bibinfo {year}
  {1992})\ pp.\ \bibinfo {pages} {177--235}\BibitemShut {NoStop}%
\bibitem [{\citenamefont {Assaad}\ and\ \citenamefont
  {Evertz}(2008)}]{Assaad2008:DQMC}%
  \BibitemOpen
  \bibfield  {author} {\bibinfo {author} {\bibfnamefont {F.F.}\ \bibnamefont
  {Assaad}}\ and\ \bibinfo {author} {\bibfnamefont {H.G.}\ \bibnamefont
  {Evertz}},\ }\enquote {\bibinfo {title} {World-line and determinantal quantum
  monte carlo methods for spins, phonons and electrons},}\ in\ \href {\doibase
  10.1007/978-3-540-74686-7_10} {\emph {\bibinfo {booktitle} {Computational
  Many-Particle Physics}}},\ \bibinfo {editor} {edited by\ \bibinfo {editor}
  {\bibfnamefont {H.}~\bibnamefont {Fehske}}, \bibinfo {editor} {\bibfnamefont
  {R.}~\bibnamefont {Schneider}}, \ and\ \bibinfo {editor} {\bibfnamefont
  {A.}~\bibnamefont {Wei{\ss}e}}}\ (\bibinfo  {publisher} {Springer Berlin
  Heidelberg},\ \bibinfo {address} {Berlin, Heidelberg},\ \bibinfo {year}
  {2008})\ pp.\ \bibinfo {pages} {277--356}\BibitemShut {NoStop}%
\bibitem [{sup()}]{suppmat}%
  \BibitemOpen
  \href@noop {} {}\bibinfo {note} {Please refer to supplementary materials for
  more details on the algorithm and the numerical simulations.}\BibitemShut
  {Stop}%
\bibitem [{\citenamefont {Feldbacher}\ and\ \citenamefont
  {Assaad}(2001)}]{Feldbacher2001:EfficientTimeDisp}%
  \BibitemOpen
  \bibfield  {author} {\bibinfo {author} {\bibfnamefont {M.}~\bibnamefont
  {Feldbacher}}\ and\ \bibinfo {author} {\bibfnamefont {F.~F.}\ \bibnamefont
  {Assaad}},\ }\bibfield  {title} {\enquote {\bibinfo {title} {Efficient
  calculation of imaginary-time-displaced correlation functions in the
  projector auxiliary-field quantum monte carlo algorithm},}\ }\href {\doibase
  10.1103/PhysRevB.63.073105} {\bibfield  {journal} {\bibinfo  {journal} {Phys.
  Rev. B}\ }\textbf {\bibinfo {volume} {63}},\ \bibinfo {pages} {073105}
  (\bibinfo {year} {2001})}\BibitemShut {NoStop}%
\bibitem [{\citenamefont {Sorella}\ \emph {et~al.}(2012)\citenamefont
  {Sorella}, \citenamefont {Otsuka},\ and\ \citenamefont
  {Yunoki}}]{Sorella2012:Absence}%
  \BibitemOpen
  \bibfield  {author} {\bibinfo {author} {\bibfnamefont {Sandro}\ \bibnamefont
  {Sorella}}, \bibinfo {author} {\bibfnamefont {Yuichi}\ \bibnamefont
  {Otsuka}}, \ and\ \bibinfo {author} {\bibfnamefont {Seiji}\ \bibnamefont
  {Yunoki}},\ }\bibfield  {title} {\enquote {\bibinfo {title} {Absence of a
  spin liquid phase in the hubbard model on the honeycomb lattice},}\ }\href
  {\doibase 10.1038/srep00992} {\bibfield  {journal} {\bibinfo  {journal}
  {Scientific Reports}\ }\textbf {\bibinfo {volume} {2}},\ \bibinfo {pages}
  {992} (\bibinfo {year} {2012})}\BibitemShut {NoStop}%
\bibitem [{\citenamefont {Assaad}\ and\ \citenamefont
  {Herbut}(2013)}]{Assaad2013:Pinning}%
  \BibitemOpen
  \bibfield  {author} {\bibinfo {author} {\bibfnamefont {Fakher~F.}\
  \bibnamefont {Assaad}}\ and\ \bibinfo {author} {\bibfnamefont {Igor~F.}\
  \bibnamefont {Herbut}},\ }\bibfield  {title} {\enquote {\bibinfo {title}
  {Pinning the order: The nature of quantum criticality in the hubbard model on
  honeycomb lattice},}\ }\href {\doibase 10.1103/PhysRevX.3.031010} {\bibfield
  {journal} {\bibinfo  {journal} {Phys. Rev. X}\ }\textbf {\bibinfo {volume}
  {3}},\ \bibinfo {pages} {031010} (\bibinfo {year} {2013})}\BibitemShut
  {NoStop}%
\bibitem [{\citenamefont {Herbut}(2006)}]{Herbut2006:Honeycomb}%
  \BibitemOpen
  \bibfield  {author} {\bibinfo {author} {\bibfnamefont {Igor~F.}\ \bibnamefont
  {Herbut}},\ }\bibfield  {title} {\enquote {\bibinfo {title} {Interactions and
  phase transitions on graphene's honeycomb lattice},}\ }\href {\doibase
  10.1103/PhysRevLett.97.146401} {\bibfield  {journal} {\bibinfo  {journal}
  {Phys. Rev. Lett.}\ }\textbf {\bibinfo {volume} {97}},\ \bibinfo {pages}
  {146401} (\bibinfo {year} {2006})}\BibitemShut {NoStop}%
\bibitem [{\citenamefont {Helmes}\ \emph {et~al.}(2016)\citenamefont {Helmes},
  \citenamefont {Hayward~Sierens}, \citenamefont {Chandran}, \citenamefont
  {Witczak-Krempa},\ and\ \citenamefont {Melko}}]{Helmes2016:UniversalCorner}%
  \BibitemOpen
  \bibfield  {author} {\bibinfo {author} {\bibfnamefont {Johannes}\
  \bibnamefont {Helmes}}, \bibinfo {author} {\bibfnamefont {Lauren~E.}\
  \bibnamefont {Hayward~Sierens}}, \bibinfo {author} {\bibfnamefont {Anushya}\
  \bibnamefont {Chandran}}, \bibinfo {author} {\bibfnamefont {William}\
  \bibnamefont {Witczak-Krempa}}, \ and\ \bibinfo {author} {\bibfnamefont
  {Roger~G.}\ \bibnamefont {Melko}},\ }\bibfield  {title} {\enquote {\bibinfo
  {title} {Universal corner entanglement of dirac fermions and gapless bosons
  from the continuum to the lattice},}\ }\href {\doibase
  10.1103/PhysRevB.94.125142} {\bibfield  {journal} {\bibinfo  {journal} {Phys.
  Rev. B}\ }\textbf {\bibinfo {volume} {94}},\ \bibinfo {pages} {125142}
  (\bibinfo {year} {2016})}\BibitemShut {NoStop}%
\bibitem [{\citenamefont {Singh}\ \emph {et~al.}(2012)\citenamefont {Singh},
  \citenamefont {Melko},\ and\ \citenamefont
  {Oitmaa}}]{Singh2012:ThermoSingEE}%
  \BibitemOpen
  \bibfield  {author} {\bibinfo {author} {\bibfnamefont {Rajiv R.~P.}\
  \bibnamefont {Singh}}, \bibinfo {author} {\bibfnamefont {Roger~G.}\
  \bibnamefont {Melko}}, \ and\ \bibinfo {author} {\bibfnamefont {Jaan}\
  \bibnamefont {Oitmaa}},\ }\bibfield  {title} {\enquote {\bibinfo {title}
  {Thermodynamic singularities in the entanglement entropy at a two-dimensional
  quantum critical point},}\ }\href {\doibase 10.1103/PhysRevB.86.075106}
  {\bibfield  {journal} {\bibinfo  {journal} {Phys. Rev. B}\ }\textbf {\bibinfo
  {volume} {86}},\ \bibinfo {pages} {075106} (\bibinfo {year}
  {2012})}\BibitemShut {NoStop}%
\bibitem [{\citenamefont {Kallin}\ \emph {et~al.}(2013)\citenamefont {Kallin},
  \citenamefont {Hyatt}, \citenamefont {Singh},\ and\ \citenamefont
  {Melko}}]{Kallin2013:EE2DQCP}%
  \BibitemOpen
  \bibfield  {author} {\bibinfo {author} {\bibfnamefont {Ann~B.}\ \bibnamefont
  {Kallin}}, \bibinfo {author} {\bibfnamefont {Katharine}\ \bibnamefont
  {Hyatt}}, \bibinfo {author} {\bibfnamefont {Rajiv R.~P.}\ \bibnamefont
  {Singh}}, \ and\ \bibinfo {author} {\bibfnamefont {Roger~G.}\ \bibnamefont
  {Melko}},\ }\bibfield  {title} {\enquote {\bibinfo {title} {Entanglement at a
  two-dimensional quantum critical point: A numerical linked-cluster expansion
  study},}\ }\href {\doibase 10.1103/PhysRevLett.110.135702} {\bibfield
  {journal} {\bibinfo  {journal} {Phys. Rev. Lett.}\ }\textbf {\bibinfo
  {volume} {110}},\ \bibinfo {pages} {135702} (\bibinfo {year}
  {2013})}\BibitemShut {NoStop}%
\bibitem [{\citenamefont {Kallin}\ \emph {et~al.}(2014)\citenamefont {Kallin},
  \citenamefont {Stoudenmire}, \citenamefont {Fendley}, \citenamefont {Singh},\
  and\ \citenamefont {Melko}}]{Kallin2014:CornerO3}%
  \BibitemOpen
  \bibfield  {author} {\bibinfo {author} {\bibfnamefont {A~B}\ \bibnamefont
  {Kallin}}, \bibinfo {author} {\bibfnamefont {E~M}\ \bibnamefont
  {Stoudenmire}}, \bibinfo {author} {\bibfnamefont {P}~\bibnamefont {Fendley}},
  \bibinfo {author} {\bibfnamefont {R~R~P}\ \bibnamefont {Singh}}, \ and\
  \bibinfo {author} {\bibfnamefont {R~G}\ \bibnamefont {Melko}},\ }\bibfield
  {title} {\enquote {\bibinfo {title} {Corner contribution to the entanglement
  entropy of an o(3) quantum critical point in 2 + 1 dimensions},}\ }\href
  {\doibase 10.1088/1742-5468/2014/06/P06009} {\bibfield  {journal} {\bibinfo
  {journal} {Journal of Statistical Mechanics: Theory and Experiment}\ }\textbf
  {\bibinfo {volume} {2014}},\ \bibinfo {pages} {P06009} (\bibinfo {year}
  {2014})}\BibitemShut {NoStop}%
\bibitem [{\citenamefont {Helmes}\ and\ \citenamefont
  {Wessel}(2014)}]{Helmes2014:EEbilayer}%
  \BibitemOpen
  \bibfield  {author} {\bibinfo {author} {\bibfnamefont {Johannes}\
  \bibnamefont {Helmes}}\ and\ \bibinfo {author} {\bibfnamefont {Stefan}\
  \bibnamefont {Wessel}},\ }\bibfield  {title} {\enquote {\bibinfo {title}
  {Entanglement entropy scaling in the bilayer heisenberg spin system},}\
  }\href {\doibase 10.1103/PhysRevB.89.245120} {\bibfield  {journal} {\bibinfo
  {journal} {Phys. Rev. B}\ }\textbf {\bibinfo {volume} {89}},\ \bibinfo
  {pages} {245120} (\bibinfo {year} {2014})}\BibitemShut {NoStop}%
\bibitem [{\citenamefont {Stoudenmire}\ \emph {et~al.}(2014)\citenamefont
  {Stoudenmire}, \citenamefont {Gustainis}, \citenamefont {Johal},
  \citenamefont {Wessel},\ and\ \citenamefont
  {Melko}}]{Stoudenmire2014:CornerEEO2}%
  \BibitemOpen
  \bibfield  {author} {\bibinfo {author} {\bibfnamefont {E.~M.}\ \bibnamefont
  {Stoudenmire}}, \bibinfo {author} {\bibfnamefont {Peter}\ \bibnamefont
  {Gustainis}}, \bibinfo {author} {\bibfnamefont {Ravi}\ \bibnamefont {Johal}},
  \bibinfo {author} {\bibfnamefont {Stefan}\ \bibnamefont {Wessel}}, \ and\
  \bibinfo {author} {\bibfnamefont {Roger~G.}\ \bibnamefont {Melko}},\
  }\bibfield  {title} {\enquote {\bibinfo {title} {Corner contribution to the
  entanglement entropy of strongly interacting o(2) quantum critical systems in
  2+1 dimensions},}\ }\href {\doibase 10.1103/PhysRevB.90.235106} {\bibfield
  {journal} {\bibinfo  {journal} {Phys. Rev. B}\ }\textbf {\bibinfo {volume}
  {90}},\ \bibinfo {pages} {235106} (\bibinfo {year} {2014})}\BibitemShut
  {NoStop}%
\bibitem [{\citenamefont {Helmes}\ and\ \citenamefont
  {Wessel}(2015)}]{Helmes2015:EEXY}%
  \BibitemOpen
  \bibfield  {author} {\bibinfo {author} {\bibfnamefont {Johannes}\
  \bibnamefont {Helmes}}\ and\ \bibinfo {author} {\bibfnamefont {Stefan}\
  \bibnamefont {Wessel}},\ }\bibfield  {title} {\enquote {\bibinfo {title}
  {Correlations and entanglement in quantum critical bilayer and necklace xy
  models},}\ }\href {\doibase 10.1103/PhysRevB.92.125120} {\bibfield  {journal}
  {\bibinfo  {journal} {Phys. Rev. B}\ }\textbf {\bibinfo {volume} {92}},\
  \bibinfo {pages} {125120} (\bibinfo {year} {2015})}\BibitemShut {NoStop}%
\bibitem [{\citenamefont {Bueno}\ \emph {et~al.}(2015)\citenamefont {Bueno},
  \citenamefont {Myers},\ and\ \citenamefont
  {Witczak-Krempa}}]{Bueno2015:UniversalCorner}%
  \BibitemOpen
  \bibfield  {author} {\bibinfo {author} {\bibfnamefont {Pablo}\ \bibnamefont
  {Bueno}}, \bibinfo {author} {\bibfnamefont {Robert~C.}\ \bibnamefont
  {Myers}}, \ and\ \bibinfo {author} {\bibfnamefont {William}\ \bibnamefont
  {Witczak-Krempa}},\ }\bibfield  {title} {\enquote {\bibinfo {title}
  {Universality of corner entanglement in conformal field theories},}\ }\href
  {\doibase 10.1103/PhysRevLett.115.021602} {\bibfield  {journal} {\bibinfo
  {journal} {Phys. Rev. Lett.}\ }\textbf {\bibinfo {volume} {115}},\ \bibinfo
  {pages} {021602} (\bibinfo {year} {2015})}\BibitemShut {NoStop}%
\bibitem [{\citenamefont {Bueno}\ and\ \citenamefont
  {Myers}(2015)}]{Bueno2015:CornerContributions}%
  \BibitemOpen
  \bibfield  {author} {\bibinfo {author} {\bibfnamefont {Pablo}\ \bibnamefont
  {Bueno}}\ and\ \bibinfo {author} {\bibfnamefont {Robert~C.}\ \bibnamefont
  {Myers}},\ }\bibfield  {title} {\enquote {\bibinfo {title} {Corner
  contributions to holographic entanglement entropy},}\ }\href {\doibase
  10.1007/JHEP08(2015)068} {\bibfield  {journal} {\bibinfo  {journal} {Journal
  of High Energy Physics}\ }\textbf {\bibinfo {volume} {2015}},\ \bibinfo
  {pages} {68} (\bibinfo {year} {2015})}\BibitemShut {NoStop}%
\bibitem [{\citenamefont {Assaad}(1999)}]{Assaad1999:SU2Invariant}%
  \BibitemOpen
  \bibfield  {author} {\bibinfo {author} {\bibfnamefont {F.~F.}\ \bibnamefont
  {Assaad}},\ }\bibfield  {title} {\enquote {\bibinfo {title} {Su(2)-spin
  invariant auxiliary field quantum monte-carlo algorithm for hubbard
  models},}\ }in\ \href@noop {} {\emph {\bibinfo {booktitle} {High Performance
  Computing in Science and Engineering '98}}},\ \bibinfo {editor} {edited by\
  \bibinfo {editor} {\bibfnamefont {Egon}\ \bibnamefont {Krause}}\ and\
  \bibinfo {editor} {\bibfnamefont {Willi}\ \bibnamefont {J{\"a}ger}}}\
  (\bibinfo  {publisher} {Springer Berlin Heidelberg},\ \bibinfo {address}
  {Berlin, Heidelberg},\ \bibinfo {year} {1999})\ pp.\ \bibinfo {pages}
  {105--111}\BibitemShut {NoStop}%
\bibitem [{wwk()}]{wwkprivate}%
  \BibitemOpen
  \href@noop {} {}\bibinfo {note} {William Witczak-Krempa, private
  communication.}\BibitemShut {Stop}%
\bibitem [{\citenamefont {Da~Liao}\ \emph {et~al.}(2023)\citenamefont
  {Da~Liao}, \citenamefont {Pan}, \citenamefont {Jiang}, \citenamefont {Qi},\
  and\ \citenamefont {Meng}}]{DaLiao:2023pdn}%
  \BibitemOpen
  \bibfield  {author} {\bibinfo {author} {\bibfnamefont {Yuan}\ \bibnamefont
  {Da~Liao}}, \bibinfo {author} {\bibfnamefont {Gaopei}\ \bibnamefont {Pan}},
  \bibinfo {author} {\bibfnamefont {Weilun}\ \bibnamefont {Jiang}}, \bibinfo
  {author} {\bibfnamefont {Yang}\ \bibnamefont {Qi}}, \ and\ \bibinfo {author}
  {\bibfnamefont {Zi~Yang}\ \bibnamefont {Meng}},\ }\bibfield  {title}
  {\enquote {\bibinfo {title} {{The teaching from entanglement: 2D SU(2)
  antiferromagnet to valence bond solid deconfined quantum critical points are
  not conformal}},}\ }\href@noop {} {\  (\bibinfo {year} {2023})},\ \Eprint
  {http://arxiv.org/abs/2302.11742} {arXiv:2302.11742 [cond-mat.str-el]}
  \BibitemShut {NoStop}%
\bibitem [{\citenamefont {Pan}\ \emph {et~al.}(2023)\citenamefont {Pan},
  \citenamefont {Da~Liao}, \citenamefont {Jiang}, \citenamefont {D'Emidio},
  \citenamefont {Qi},\ and\ \citenamefont {Meng}}]{Pan:2023ysg}%
  \BibitemOpen
  \bibfield  {author} {\bibinfo {author} {\bibfnamefont {Gaopei}\ \bibnamefont
  {Pan}}, \bibinfo {author} {\bibfnamefont {Yuan}\ \bibnamefont {Da~Liao}},
  \bibinfo {author} {\bibfnamefont {Weilun}\ \bibnamefont {Jiang}}, \bibinfo
  {author} {\bibfnamefont {Jonathan}\ \bibnamefont {D'Emidio}}, \bibinfo
  {author} {\bibfnamefont {Yang}\ \bibnamefont {Qi}}, \ and\ \bibinfo {author}
  {\bibfnamefont {Zi~Yang}\ \bibnamefont {Meng}},\ }\bibfield  {title}
  {\enquote {\bibinfo {title} {{Stable computation of entanglement entropy for
  two-dimensional interacting fermion systems}},}\ }\href {\doibase
  10.1103/PhysRevB.108.L081123} {\bibfield  {journal} {\bibinfo  {journal}
  {Phys. Rev. B}\ }\textbf {\bibinfo {volume} {108}},\ \bibinfo {pages}
  {L081123} (\bibinfo {year} {2023})},\ \Eprint
  {http://arxiv.org/abs/2303.14326} {arXiv:2303.14326 [cond-mat.str-el]}
  \BibitemShut {NoStop}%
\bibitem [{\citenamefont {Liu}\ \emph {et~al.}(2023{\natexlab{b}})\citenamefont
  {Liu}, \citenamefont {Liao}, \citenamefont {Pan}, \citenamefont {Song},
  \citenamefont {Zhao}, \citenamefont {Jiang}, \citenamefont {Jian},
  \citenamefont {You}, \citenamefont {Assaad}, \citenamefont {Meng},\ and\
  \citenamefont {Xu}}]{Liu:2023msa}%
  \BibitemOpen
  \bibfield  {author} {\bibinfo {author} {\bibfnamefont {Zi~Hong}\ \bibnamefont
  {Liu}}, \bibinfo {author} {\bibfnamefont {Yuan~Da}\ \bibnamefont {Liao}},
  \bibinfo {author} {\bibfnamefont {Gaopei}\ \bibnamefont {Pan}}, \bibinfo
  {author} {\bibfnamefont {Menghan}\ \bibnamefont {Song}}, \bibinfo {author}
  {\bibfnamefont {Jiarui}\ \bibnamefont {Zhao}}, \bibinfo {author}
  {\bibfnamefont {Weilun}\ \bibnamefont {Jiang}}, \bibinfo {author}
  {\bibfnamefont {Chao-Ming}\ \bibnamefont {Jian}}, \bibinfo {author}
  {\bibfnamefont {Yi-Zhuang}\ \bibnamefont {You}}, \bibinfo {author}
  {\bibfnamefont {Fakher~F.}\ \bibnamefont {Assaad}}, \bibinfo {author}
  {\bibfnamefont {Zi~Yang}\ \bibnamefont {Meng}}, \ and\ \bibinfo {author}
  {\bibfnamefont {Cenke}\ \bibnamefont {Xu}},\ }\bibfield  {title} {\enquote
  {\bibinfo {title} {{Disorder Operator and R\'enyi Entanglement Entropy of
  Symmetric Mass Generation}},}\ }\href@noop {} {\  (\bibinfo {year}
  {2023}{\natexlab{b}})},\ \Eprint {http://arxiv.org/abs/2308.07380}
  {arXiv:2308.07380 [cond-mat.str-el]} \BibitemShut {NoStop}%
\bibitem [{\citenamefont {Da~Liao}(2023)}]{DaLiao:2023klr}%
  \BibitemOpen
  \bibfield  {author} {\bibinfo {author} {\bibfnamefont {Yuan}\ \bibnamefont
  {Da~Liao}},\ }\bibfield  {title} {\enquote {\bibinfo {title} {{Controllable
  Incremental Algorithm for Entanglement Entropy and Other Observables with
  Exponential Variance Explosion in Many-Body Systems}},}\ }\href@noop {} {\
  (\bibinfo {year} {2023})},\ \Eprint {http://arxiv.org/abs/2307.10602}
  {arXiv:2307.10602 [cond-mat.str-el]} \BibitemShut {NoStop}%
\end{thebibliography}%

\section{Supplemental material}

\subsection{Efficient Green's function updates}
Here we show how the Green's function at time slice $\theta$, at the center of the wavefunction overlap, is updated after flipping a HS field at time slice $\tau$.  Following the notation of [\onlinecite{Assaad2008:DQMC}], the imaginary time displaced Green's functions are given by

\begin{equation}
\label{eq:Gt1t2}
\begin{split}
G^s(\tau_2,\tau_1) &= B^s(\tau_2,\tau_1) G^s(\tau_1) \quad \quad \quad \quad \quad\quad \, \, \, \tau_2 > \tau_1 \\
G^s(\tau_2,\tau_1) &= - ( \mathds{1}-G^s(\tau_2)) B^s(\tau_1,\tau_2)^{-1} \quad\quad \tau_2 < \tau_1
\end{split}
\end{equation}
where $B^s(\tau_2,\tau_1)$ is the imaginary time evolution (the Trotter slice matrices) from $\tau_1$ to $\tau_2$ with HS field configuration $s$.  Changing the HS field at time slice $\tau$ changes the Green's function at time slice $\theta$ as follows:
\begin{equation}
\label{eq:Gtheta}
G^{\tilde{s}}(\theta) =  G^s(\theta) + \gamma_i G^s(\theta, \tau) \bm{v}_i \bm{v}^{T}_i G^s(\tau, \theta) .
\end{equation}
In the SU(2) invariant HS decomposition $\gamma_i = (1/(1-e^{-2 i \alpha \tilde{\sigma}_i})  - G_{i,i}^s(\tau) )^{-1}$ with $\alpha = \arccos(e^{-\frac{\Delta_{\tau}U}{2}})$,  and $\tilde{\sigma}_i$ is the new HS field value on site $i$ at time slice $\tau$.  $ \bm{v}_i$ is a column vector that is one at element $i$ and zero elsewhere, so $G^{\tilde{s}}(\theta)$ is obtained by a simple rank-1 update.

Changing the HS field also changes the imaginary time displaced greens functions.  These get updated as follows
\begin{equation}
\label{eq:Gt1t2new}
\begin{split}
G^{\tilde{s}}(\theta,\tau) &= G^s(\theta,\tau) + \gamma_i G^s(\theta, \tau) \bm{v}_i \bm{v}^{T}_i (\mathds{1}-G^s(\tau)) \\
G^{\tilde{s}}(\tau,\theta) &= G^s(\tau,\theta) + \gamma_i G^s(\tau) \bm{v}_i \bm{v}^{T}_i G^s(\tau,\theta).
\end{split}
\end{equation}
And finally we perform the standard update of $G^s(\tau)$
\begin{equation}
\label{eq:Gtau}
G^{\tilde{s}}(\tau) =  G^s(\tau) - \gamma_i G^s(\tau) \bm{v}_i \bm{v}^{T}_i (\mathds{1}-G^s(\tau)) .
\end{equation}

\subsection{Efficient HS field flips}

An essential ingredient of the DQMC algorithm is the efficient evaluation of the weight ratio upon flipping a HS field value.
\begin{equation}
\label{eq:Rhssup}
R=\frac{W_{s'_1}}{W_{s_1}}\frac{\det(g^{s'_1,s_2}_C)}{\det(g^{s_1,s_2}_C)}.
\end{equation}
Remarkably, the inclusion of the new Grover factor only slightly modifies the standard ratio formula.  For one spin species this is proportional to (neglecting an overall factor)
\begin{equation}
\label{eq:Reasy}
R_{\uparrow} = 1 + (e^{2 i \alpha \tilde{\sigma}_i}-1)(1 - G_{i,i}^{s_1}(\tau) - \Gamma),
\end{equation}
where all correlations between the replicas are contained in
\begin{equation}
\label{eq:Gamma}
 \Gamma \equiv G^{s_1}_{i,C}(\tau,\theta) (\mathds{1} - 2 G^{s_2}_{C}(\theta)) (g^{s_1,s_2}_C)^{-1}G^{s_1}_{C,i}(\theta,\tau) .
\end{equation}
Here the parts of the matrices are in the subscript and a single subscript means a square matrix in that region.  Provided that the Grover inverse $(g^{s_1,s_2}_C)^{-1}$ is stored and maintained, this factor is computed with $\mathcal{O}(N^2_C)$ multiplications.  This is still efficient since the standard Green's function update upon acceptance requires $\mathcal{O}(N^2_{\text{site}})$ multiplications.

\subsection{Updating the Grover inverse}

Since we need to maintain the Grover inverse to efficiently compute weight ratios, we need to know how to update it when the HS field gets flipped.  Assuming we accept the move in the previous section, the rank-1 update is given by
\begin{equation}
\label{eq:g_inv}
(g^{\tilde{s}_1,s_2}_C)^{-1} = (g^{s_1,s_2}_C)^{-1} + \rho_i \bm{a}_i \bm{b}^T_i,
\end{equation}
where
\begin{equation}
\label{eq:a}
\begin{split}
\bm{a}_i &= (g^{s_1,s_2}_C)^{-1} G^{s_1}_{C,i}(\theta,\tau) \\
\bm{b}^T_i &= G^{s_1}_{i,C}(\tau,\theta) (\mathds{1} - 2 G^{s_2}_{C}(\theta)) (g^{s_1,s_2}_C)^{-1}\\
\rho_i &= (1/(1-e^{-2 i \alpha \tilde{\sigma}_i})  - G_{i,i}^{s_1}(\tau) - \Gamma)^{-1}.
\end{split}
\end{equation}
Again, this is an efficient $\mathcal{O}(N^2_C)$ update, and one can make use of the fact that $\bm{b}^T_i$ was already evaluated when computing the ratio.

\subsection{Updating the entangling region}

If a site $i$ is in $A$ but has not yet joined the dynamical entangling region $C$, we can ask it to 
join with probability
\begin{equation}
\label{eq:pjoin}
P_{\text{join}} =\min\left\{1, \frac{\lambda}{1-\lambda} \frac{\det(g^{s_1,s_2}_{C+i})}{\det(g^{s_1,s_2}_{C})} \right\}
\end{equation}
Likewise if a site $i$ is in $C$ we can ask it to leave with probability
\begin{equation}
\label{eq:psplit}
P_{\text{split}} =\min\left\{1, \frac{1-\lambda}{\lambda} \frac{\det(g^{s_1,s_2}_{C-i})}{\det(g^{s_1,s_2}_{C})} \right\}
\end{equation}
Though not essential, we find improved performance by avoiding the calculation of Grover determinants as we will conceptually outline here.

The idea is to use the block structure of $g^{s_1,s_2}_{C+i}$, for instance, in order to express the determinant in terms of the determinant in the sub-block $C$, which can be cancelled out.  Again this improved formula involves the Grover inverse, which is held in memory, and the probability can be computed in $\mathcal{O}(N^2_C)$ multiplications.  Intermediate calculations on the blocks can then be used to obtain the enlarged Grover inverse $(g^{s_1,s_2}_{C+i})^{-1}$ without having to recompute it from scratch.  A similar procedure goes for splitting a site.

\subsection{Details on computing the equilibrium $S_2$ estimator}
Here we would like to explain in a bit more detail how the second R\'enyi entanglement entropy is computing using the formula that we have introduced in the main text:
\begin{equation}
\label{eq:suppZrat}
\frac{\mathcal{Z}(\lambda_j)}{\mathcal{Z}(\lambda_i)}=\left\langle \left(\frac{\lambda_j}{\lambda_i}\right)^{N_C} \left(\frac{1-\lambda_j}{1-\lambda_i}\right)^{N_A-N_C} \right\rangle_{\lambda_i}.
\end{equation}
In the formula the stochastic variable is $N_C$, or the size of the subset entangling region.  The Monte Carlo will stochastically sample all possible regions $C$ that are subsets of the region of interest $A$ according to their weight in the generalized ensemble (Eq. (3) in the main text).  Eq. (\ref{eq:suppZrat}) is used to compute each of the ratios appearing in $e^{-S^A_2} =\frac{Z_A}{Z_{\o}} = \frac{\mathcal{Z}(\lambda_1)}{\mathcal{Z}(0)}\frac{\mathcal{Z}(\lambda_2)}{\mathcal{Z}(\lambda_1)}...\frac{\mathcal{Z}(1)}{\mathcal{Z}(\lambda_{N_\lambda})}$.  For all factors except the first one, this formula can be straightforwardly applied, and in fact we store a histogram of the $N_C$ values encountered in all simulations and compute the ratio in post processing.  When computing the first factor $\frac{\mathcal{Z}(\lambda_1)}{\mathcal{Z}(0)}$, the formula cannot be applied as such.  Instead one can use Eq. (\ref{eq:suppZrat}) to compute the inverse of this factor $\frac{\mathcal{Z}(0)}{\mathcal{Z}(\lambda_1)}$.

\begin{figure}[!t]
\centerline{\includegraphics[angle=0,width=1.05\columnwidth]{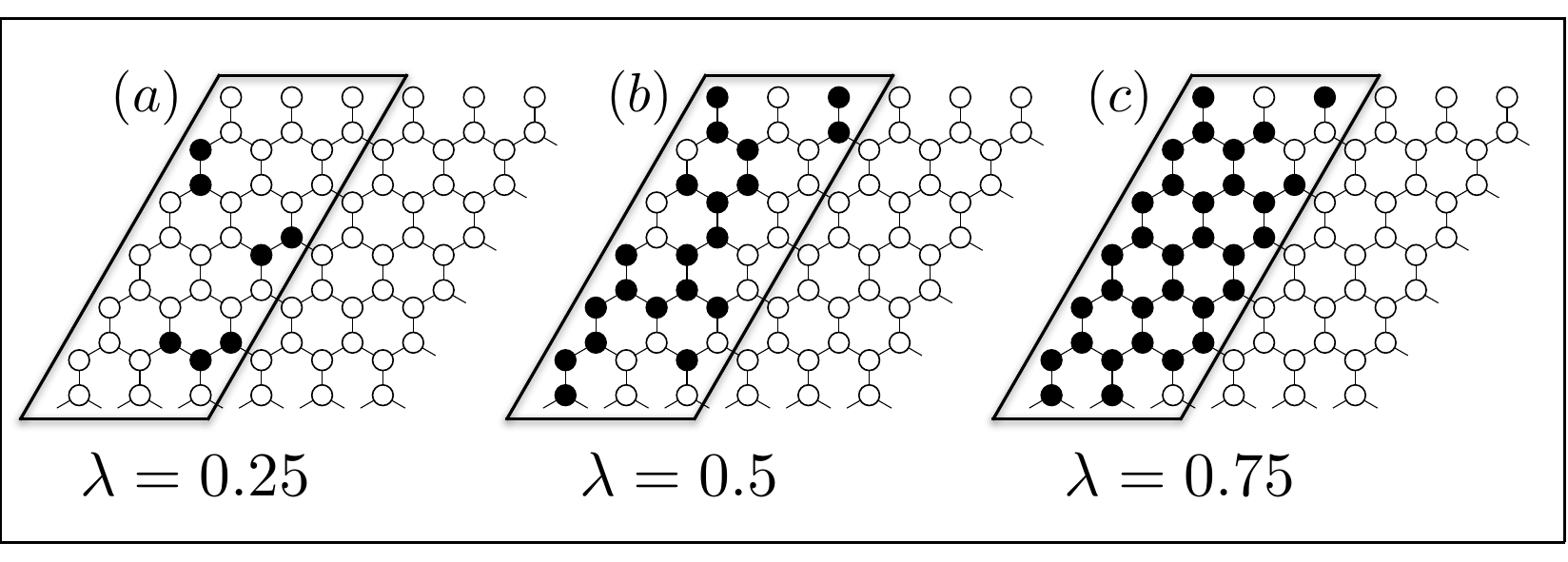}}
\caption{Pictures of typical configurations for different values of $\lambda$.  The subset entangling region $C$ is given by the black sites, whereas the entire $A$ region (the one for which we are computing the entanglement entropy) is enclosed by the box.}
\label{fig:ncconfigs}
\end{figure}

In Fig. (\ref{fig:ncconfigs}) we depict what typical configurations of the subset region $C$ look like for different values of $\lambda$.  The sites that are in the current region $C$ are shown in black, while the box encloses all sites in the region $A$.  As $\lambda$ is increased, the number of sites in the region $C (N_C)$ begins to increase, eventually populating the $A$ region entirely as $\lambda \to 1$.

\subsection{Numerical stability}
It is important to emphasize that the interaction between the two DQMC replicas only manifests itself when choosing to flip a HS spin.  Apart from this, they can be regarded as normal configurations and therefore they can be stabilized using the standard techniques.  This involves chains of QR matrix decompositions [\onlinecite{Loh1992:DQMC}] to avoid exponentially large and small values appearing in the product of Trotter slice matrices.  

However, two additional concerns are raised with this new algorithm.  Firstly, the imaginary time displaced Green's functions are always needed to flip HS spins, so they need to be maintained to high precision arbitrarily far from the center slice at $\tau = \theta$.  Secondly, in the traditional algorithm for computing $S_2$, the determinant of the Grover matrix becomes exponentially small, which signals severe ill conditioning that would prevent accurate computations of the inverse.  However we find that neither of these concerns cause problems in both the finite temperature and ground state formulations of DQMC (we have implemented both).

The first issue is especially concerning in the ground state formulation, since the imaginary time displaced Green's functions cannot be easily recomputed from scratch.  However, they can be stabilized very efficiently since the equal-time Green's function is a projector [\onlinecite{Feldbacher2001:EfficientTimeDisp}].  We therefore center our Monte Carlo sweeps around $\tau = \theta$ (the center of the wavefunction overlap) and step outwards.  This way each sweep begins with a fresh calculation of $G^s(\tau,\theta)$ and  $G^s(\theta,\tau)$ (for both replicas), and these are periodically stabilized while stepping outwards according to the prescription in [\onlinecite{Feldbacher2001:EfficientTimeDisp}], while $G^s(\tau)$ and $G^s(\theta)$ are periodically recomputed from scratch using intermediate matrix decompositions.  Since $G^s(\theta)$ is always updated with $G^s(\tau,\theta)$ and  $G^s(\theta,\tau)$ using Eq. (\ref{eq:Gtheta}), the difference between the updated and recomputed $G^s(\theta)$ tells us how well the imaginary time displaced Green's functions are maintained.  

At $U=8$, where numerical precision suffered the most, we have the following average precisions ($L_x=18,L_y=9$): $\Delta G(\tau),\Delta G(\theta) \approx 10^{-13},10^{-11}$ at $\lambda=0.0006$ where the replicas are essentially independent, $\Delta G(\tau),\Delta G(\theta) \approx 10^{-8},10^{-7}$ at $\lambda=0.5$, and $\Delta G(\tau),\Delta G(\theta) \approx 10^{-8},10^{-7}$ at $\lambda=0.9994$ where the half system is essentially fully joined.  We stress that these are the worst average precisions (absolute value of the difference between matrix elements of the current and recomputed Green's functions), which are still highly precise.  Also we observe little dependence on the system size.

The second concern regarding the ill-conditioning of the Grover matrix does not arise.  Heuristically, since $\det(g^{s_1,s_2}_C)$ appears in the DQMC weight, the nearly singular configurations that appear with uncorrelated replicas are suppressed in this formulation.  One of the most relevant tests for numerical stability is to monitor the average complex part of the acceptance ratios using the SU(2) invariant HS decomposition.  Here we find the average complex part divided by $\min(1,\text{real part})$ is never larger than $\approx 10^{-6}$ and is generally orders of magnitude smaller than this.

\begin{figure}[!t]
\centerline{\includegraphics[angle=0,width=1.05\columnwidth]{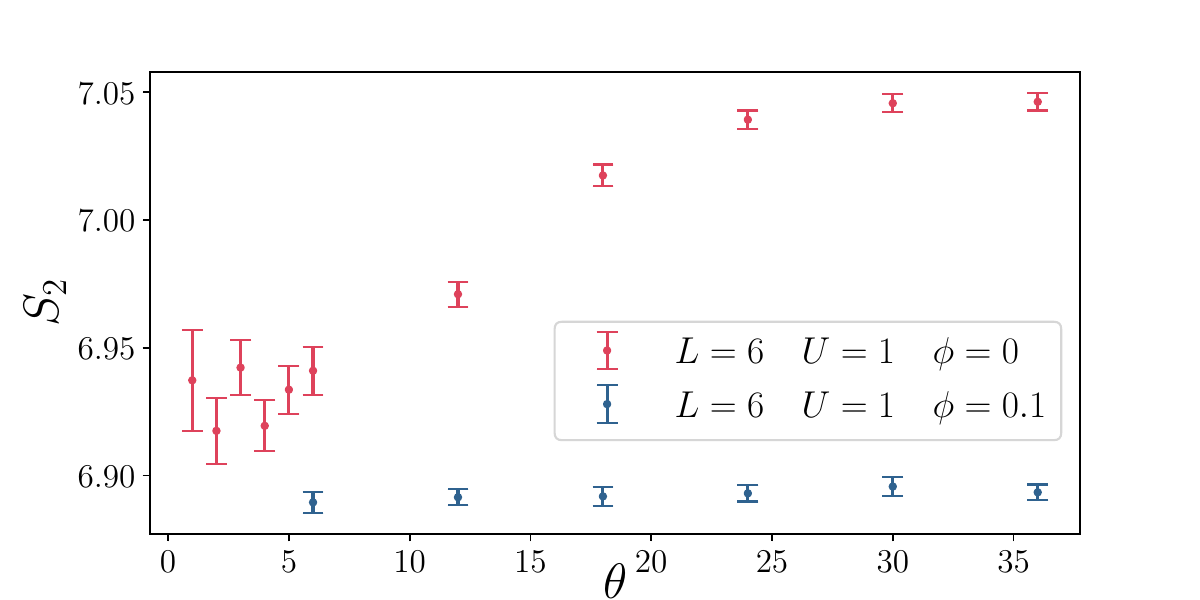}}
\caption{$U=1$ ground state convergence with and without twisted boundary conditions. Here we use zigzag triangles and $\Delta_{\tau}=0.1$}
\label{fig:convergephi}
\end{figure}

\begin{figure}[!t]
\centerline{\includegraphics[angle=0,width=1.05\columnwidth]{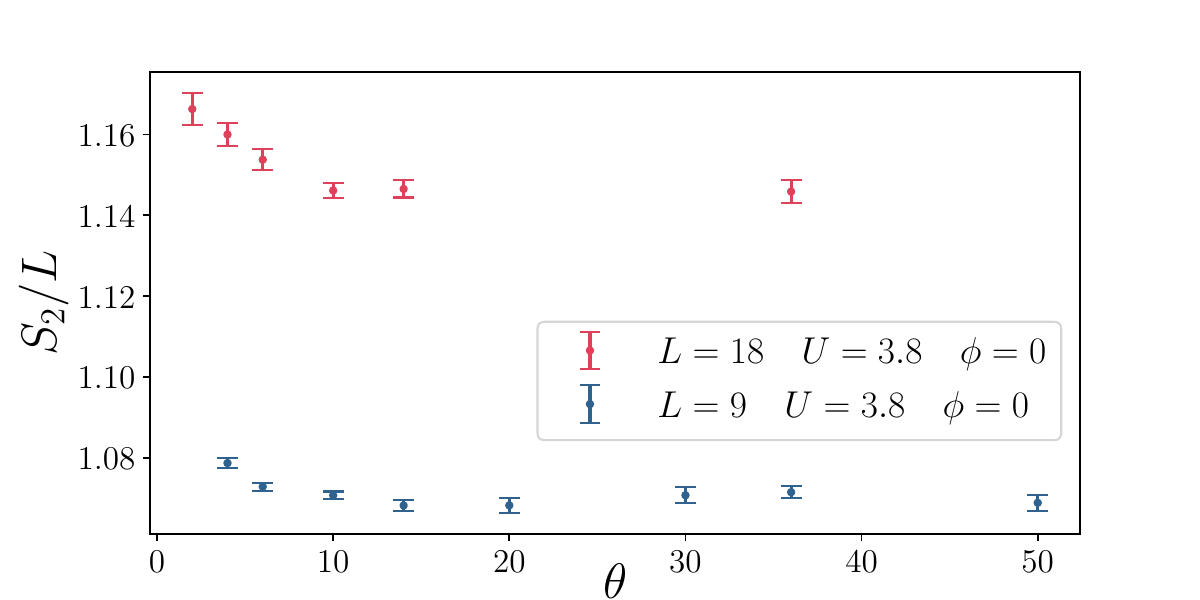}}
\caption{Zigzag triangle convergence at $U=3.8$ as a function of $\theta$ for two system sizes using no twist ($\phi=0$) and  $\Delta_{\tau}=0.1$.}
\label{fig:converge}
\end{figure}

\begin{figure}[!h]
\centerline{\includegraphics[angle=0,width=1.05\columnwidth]{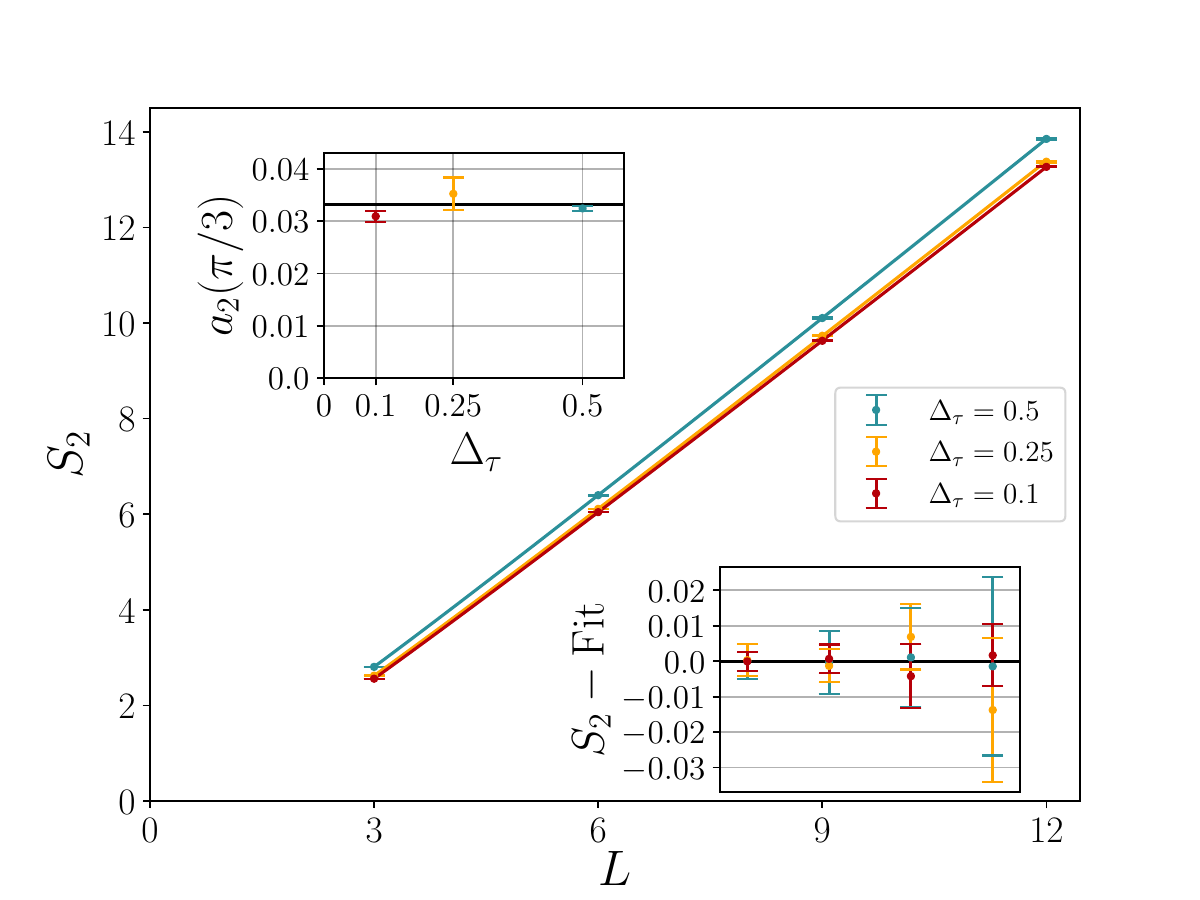}}
\caption{The Trotter dependence of our data at the critical point $U=3.8$ with the triangular regions in the main text.  At $\Delta_{\tau}=0.5$ there is a noticeable Trotter error, however this does not affect the fit to the universal corner piece, as shown by the extracted value of $a_2(\pi/3)$ in the upper inset.  The black line in the inset is the field theory value for a single Dirac fermion.  Even on these small system sizes, we observe very good fits to the scaling form, which we show in the inset on the lower right.}
\label{fig:dtstudy}
\end{figure}

\begin{figure}[!h]
\centerline{\includegraphics[angle=0,width=1.05\columnwidth]{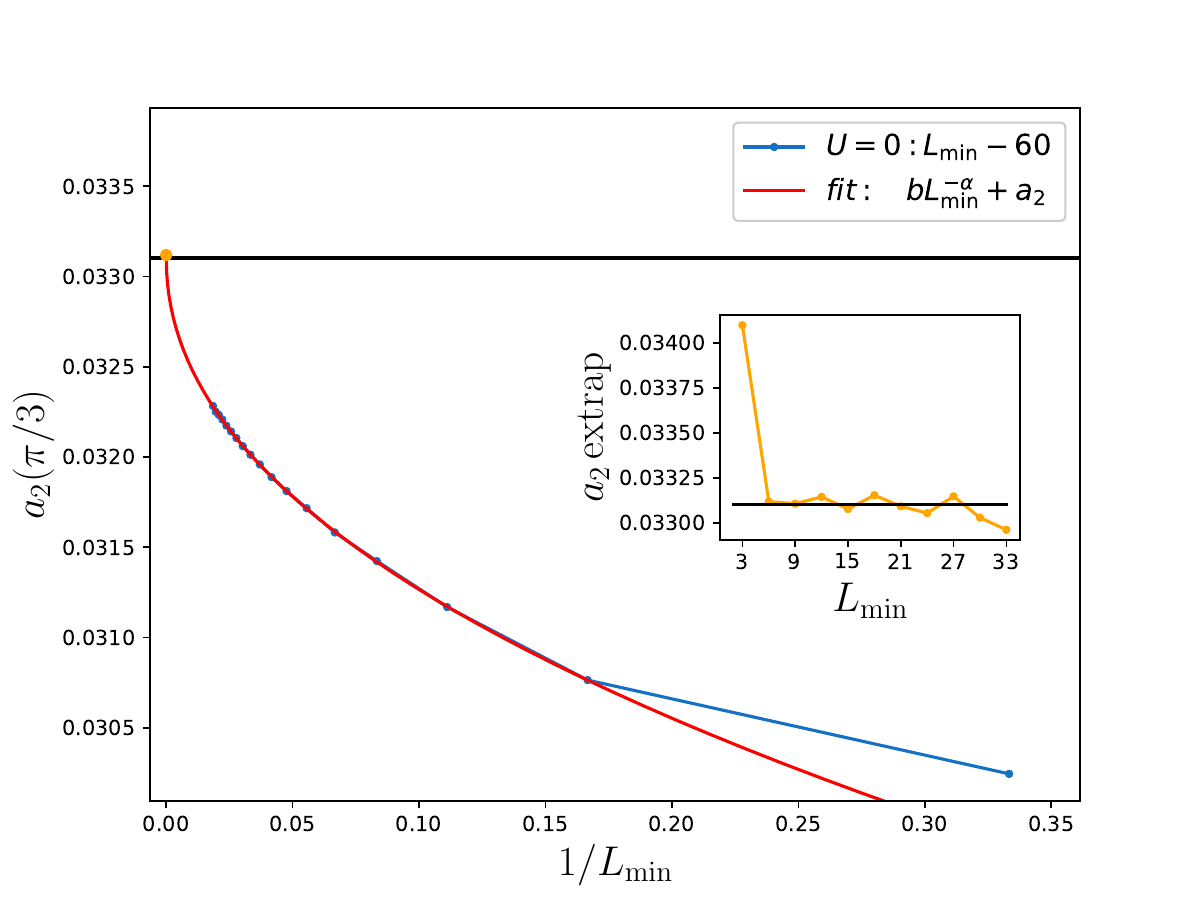}}
\caption{The fitted corner coefficients using a fit window $L_{\text{min}}$ to $L_{\text{max}}=60$.  We have extrapolated these values to $L_{\text{min}} \to \infty$ using a power law fit of all the points greater than or equal to $L_{\text{min}}$.  The example fit in red uses $L_{\text{min}}=6$ with the extrapolated value given as the orange point. The extrapolated values are given in the inset as a function of $L_{\text{min}}$.  Even using a fixed $L_{\text{max}}=60$, we find remarkably agreement with the field theoretic prediction for a free Dirac fermion.}
\label{fig:U0a2extrap}
\end{figure}

\subsection{Ground state convergence}

Here we provide additional data showing convergence to the ground state, with special attention paid to the semi-metal phase, which we found to be particularly challenging.  Fig. (\ref{fig:convergephi}) shows the ground state convergence at $U=1$ for an $L=6$ zigzag triangle with $\Delta_{\tau}=0.1$.  Here the red points show the result of evolving the trial wave function (obtained by diagonalizing the free Hamiltonian with a very small flux threading in order to lift the degeneracy) according to the HS decomposed Hamiltonian with no flux.  We see in this case very long convergence times before the ground state value of $S_2$ is reached.  On the other hand, if we work with a finite flux (twisted boundary conditions), we observe rapid convergence to the ground state value as shown by the blue points.  Specifically, we choose the trial wave function as the ground state of the noninteracting model with flux threading, and evolve this state with the HS decomposed interacting Hamiltonian with the same flux.

We found that ground state convergence away from the semi-metal phase is easier, so typically we do not use twisted boundary conditions here.  This is shown in Fig. (\ref{fig:converge}) for different size zigzag triangles at the critical point $U=3.8$ using $\Delta_{\tau}=0.1$.

\subsection{Dependence on trotter step}

We have also studied the dependence of universal terms in the entanglement entropy on the Trotter step.  This is illustrated in Fig. (\ref{fig:dtstudy}), where we have used three different Trotter steps with our four smallest system sizes at the critical point $U=3.8$ for zigzag triangles.  We observe a large Trotter error in our raw data at $\Delta_{\tau}=0.5$, however the numerical fit gives an extracted corner contribution that is insensitive to the Trotter error.  This extra data further supports the notion that the zigzag triangle produces a corner log corresponding to free fermions, and is not enhanced at the critical point.

\subsection{Additional algorithmic details and values of $N_{\lambda}$ used}
We use a checkerboard breakup of the kinetic term in the Hamiltonian for efficiency and the trial state is taken as the ground state of the free Hamiltonian with infinitesimal flux threaded through the torus to lift the degeneracy at the Dirac point.  The symmetric Trotter decomposition is used and the $SU(2)$ symmetric HS decomposition is used.

As mentioned in the main text, we must increase the number of $\lambda$ values used as our lattices become larger.  Like any reweighting technique, we require good overlap of our histograms between adjacent values of $\lambda$.  The stochastic variable here is the number of sites in the entangling region ($N_C$), so histograms of $N_C$ should have good overlap.  We have not performed detailed studies of the required number of different $\lambda$ values in order to ensure statistically converged averages of $S_2$, but rather we use an overabundance of caution and set this number quite high.  For the triangular regions in the main text we typically used $N_{\lambda}=8,16,24,32,48,64$ for system sizes $L=3,6,9,12,15,18$ respectively.  With so many $\lambda$ values for these relatively small system sizes, the histograms between adjacent values deviate only very slightly, so we believe there to be no issue along these lines.

We did observe that using a linear ramp $\lambda_i = i/(N_{\lambda}+1)$ produces a much larger error bar for the endpoints $1/(N_{\lambda}+1)$ and $N_{\lambda}/(N_{\lambda}+1)$ as compared to contributions from interior points.  In order to avoid this we instead choose a smooth ramp given by $\lambda_i \to \sin^2(\pi \lambda_i / 2)$, where this new grid of $\lambda$ points shows stochastic error contributions that taper off near the edges.  We have checked that the overall error bar is at least the same if not better with the smooth grid.

\subsection{Corner coefficient at $U=0$}

Here we show an analysis of data that we have computed in the free case ($U=0$) using the zigzag triangle.  We wish to gauge the accuracy of the corner coefficient upon extrapolation to the thermodynamic limit.  To perform this analysis we first compute $S_2$ with zigzag triangles for system sizes $L=3,6,9,...,60$.  Next we perform a three parameter fit to a linear plus log form including system sizes $L_{\text{min}},L_{\text{min}}+3,L_{\text{min}}+6,...,60$ and extract the logarithmic coefficient, which gives us an estimate of the corner term $a_2(\pi / 3)$.  We then do this process many times for different values of $L_{\text{min}}$, which is shown as the blue points in Fig. (\ref{fig:U0a2extrap}).  These points show a very nice looking extrapolation to the true value $a_2(\pi / 3) \approx 0.03310$ [\onlinecite{Helmes2016:UniversalCorner}], so we now use the blue points to fit to a power law scaling form that we can use to extrapolate to $1/L_{\text{min}} = 0$.  This is shown as the red line, where the smallest $L_{\text{min}}$ value used in the fit was $L_{\text{min}}=6$ in this case.  This gives an extrapolated value shown by the orange dot.  Now this process can be done many times, each time excluding more small values of $L_{\text{min}}$ from the power law fit.  The resulting extrapolated points are shown in the inset, where we find extremely precise agreement.

\end{document}